\documentclass[aps,prl,twocolumn,superscriptaddress,floatfix, 10pt,longbibliography, nofootinbib]{revtex4-2}
\usepackage{graphicx}
\renewcommand{\selectlanguage}[1]{}
\usepackage{verbatim}
\usepackage{lineno}
\usepackage{bm}
\usepackage{times}
\usepackage{subfigure}
\usepackage{amsmath}
\usepackage{physics}
\usepackage{slashed}
\usepackage{subfiles}
\usepackage{lipsum}
\usepackage[normalem]{ulem}
\makeatletter

\usepackage{titlesec}

\usepackage[colorlinks,linkcolor=blue,anchorcolor=blue, citecolor=blue,urlcolor=blue]{hyperref}

\titleformat*{\section}{\large\centering\bfseries}

\usepackage[usenames,dvipsnames,svgnames,table]{xcolor}
\definecolor{BrightPurple}{RGB}{190,0,200}

\newcommand{\ASc}[1]{{\color{blue} [AScomment: #1]}}

\begin{document}

\title{
%Plasmon Damping Near Deconfined Topological Phase Transitions \\
%Or \\
%Electromagnetic Responses across Phase Transitions at the Half-Filled Chern Band
%Electromagnetic Response at a Continuous Phase Transitions in a Half-Filled Chern Band 
%Electromagnetic Response of a Half-Filled Chern Band Near Criticality
%Electromagnetic Response of a Half-Filled Chern Band Near a Continuous Phase Transition
Electromagnetic Response of a Half-Filled Chern Band near Topological Criticality
}

\author{Xinlei Yue}
\affiliation{Department of Condensed Matter Physics, Weizmann Institute of Science, Rehovot, Israel 7610001}
\author{Fabian Pichler}
\affiliation{Technical University of Munich, TUM School of Natural Sciences, Physics Department, 85748 Garching, Germany}
\affiliation{Munich Center for Quantum Science and Technology (MCQST), Schellingstr. 4, 80799 München, Germany}
\author{Michael Knap}
\affiliation{Technical University of Munich, TUM School of Natural Sciences, Physics Department, 85748 Garching, Germany}
\affiliation{Munich Center for Quantum Science and Technology (MCQST), Schellingstr. 4, 80799 München, Germany}
\author{Ady Stern}
\affiliation{Department of Condensed Matter Physics, Weizmann Institute of Science, Rehovot, Israel 7610001}
\begin{abstract}
We evaluate electromagnetic-response observables in a half-filled Chern band, across a topological phase transition between a composite Fermi liquid (CFL) and a Fermi liquid (FL) phase. While a sharp gapped plasma mode exists deep in the CFL phase, we demonstrate that it is damped near the proposed continuous phase transition between CFL and FL. This plasmon-damping phenomenon originates from emergent gauge fields and a Dirac-fermion-like spectrum. Similar features also occur in other continuous deconfined topological phase transitions, such as the Laughlin to superfluid transition in a bosonic system. In particular, this damping behavior extends over a finite range across the phase boundary, and, hence, we expect it to persist even when the transition is weakly first-order. Furthermore, we analyze the characteristic conductivity behavior, such as the Drude weight, the wavevector-dependent conductivity,  the chiral mirror effect, and thermodynamic behavior, including compressibility and magnetic susceptibility across these topological phase transitions.
\end{abstract}
\maketitle

\textit{Introduction}.---Striking experimental developments in two-dimensional materials have opened new avenues for realizing phases with topological order. Notably, fractional Chern insulators (FCIs)~\cite{Hafezi_fractional_2007, kapit_exact_2010, neupert_fractional_2011, qi_generic_2011, regnault_fractional_2011, tang_high-temperature_2011,wang_nearly_2011,murthy_hamiltonian_2012, barkeshli_topological_2012, wu_bloch_2013,moller_fractional_2015,hu_hyperdeterminants_2024} have been discovered at zero magnetic field in twisted MoTe$_2$~\cite{cai_signatures_2023,park_observation_2023,xu_observation_2023,zeng_thermodynamic_2023, xu_signatures_2025, liu_fractional_2025} and rhombohedral graphene~\cite{lu_fractional_2024,han_large_2024,han_signatures_2025,lu_extended_2025}, providing concrete platforms for studying strongly correlated phases in Chern bands. %MoTe2 FQH
The exceptional tunability and control of these systems have enabled the exploration of rich phase diagrams and sparked theoretical interest in continuous topological transitions between fractionalized states~\cite{wen_transitions_1993,chen_mott_1993,ludwig_integer_1994,ye_coulomb_1998,wen_continuous_2000,barkeshli_continuous_2012,grover_quantum_2013,barkeshli_continuous_2014,grover_entanglement_2014,barkeshli_continuous_2015,lee_emergent_2018,ma_emergent_2020,song_deconfined_2023,song_phase_2024,han_exotic_2024,kuhlenkamp_chiral_2024,wu_bipartite_2025, divic_chiral_2025, divic_anyon_2025, lu_continuous_2025,lotric_paired_2025,wang_emergent_2025,zhou_chern-simons-matter_2025,yahuizhang_continuous_2025,zhang_continuous_2025,pichler_microscopic_2025, kuhlenkamp_robust_2025, chen_topological_2025, zhang_pathways_2025}. % compared to heterostructure systems~\cite{prange_quantum_1990} 
% Therefore, many novel phase transitions are observed, among which are phase transitions involving fractionalized states at zero magnetic field. 
A particularly intriguing example is the transition between a composite Fermi liquid (CFL)~\cite{halperin_theory_1993} and a Fermi liquid (FL), taking place in a half-filled Chern band as the ratio of interaction strength to band width is varied~\cite{abouelkomsan_quantum_2023,dong_composite_2023, Goldman_zerofield_2023,abouelkomsan_compressible_2025}. Recent experiments~\cite{park_observation_2023, lu_extended_2025} have already observed the CFL and FL phases nearby in parameter space, although at low temperature they seem to be separated by an intermediate phase, potentially an insulating state or extended quantum Hall state. Theoretically, the CFL to FL phase transition has attracted further interest since it is proposed to be possibly a deconfined topological phase transition (TPT) beyond the Landau phase transition paradigm~\cite{barkeshli_continuous_2012, song_phase_2024}. Numerical works~\cite{barkeshli_continuous_2015,lu_continuous_2025,lotric_paired_2025,wang_emergent_2025} 
have provided strong evidence for a continuous deconfined TPT between a bosonic Laughlin state and a superfluid phase, which is closely related to the CFL to FL transition in the deconfined TPT framework. However, to develop a better experimental understanding of these exotic phase transitions, it is pertinent to identify their unique signatures.
%suggest the deconfined TPT to be possible via identifying a continuous phase transition between the bosonic Laughlin state and the superfluid state, 
%Experimentally, however, the CFL to FL transition is suggested to be first order by the semicircle trajectory of the resistance and conductance tensor along the phase transition. While the presence of a first-order phase transition does not rule out the possibility of a deconfined TPT, it leaves the experimental examination a hard job, as one cannot access the critical properties. 

%One may ask if there are other observables that support a deconfined TPT scenario. 
In this work, we examine the evolution of the electromagnetic response across a transition between a CFL and an FL, focusing on identifying characteristics of the two phases. These characteristics include the plasma mode, the Drude weight, the conductivity at non-zero wave vector, chiral mirroring, compressibility, and magnetic susceptibility. 
%Here, we show that the plasmon mode is damped near the deconfined TPT between CFL and FL. This behavior roots 
Their evolution is controlled by two key ingredients of deconfined TPTs: emergent gauge fields and a Dirac fermion spectrum. We analyze some of these signatures also in other deconfined TPTs, including the Laughlin-superfluid transition~\cite{barkeshli_continuous_2014} and the Laughlin-insulator-superfluid (CFL-insulator*-FL) transition~\cite{barkeshli_continuous_2012,barkeshli_continuous_2014}. While we assume a continuous transition in our calculations, we expect qualitatively similar behavior to appear even if the deconfined TPTs become weakly first-order.
Thereby, we establish unique signatures that can be probed experimentally using %quickly developing 
THz spectroscopy; see e.g. Refs.~\cite{xu_electronic_2024,chen_direct_2025, chen_terahertz_2025, pierce_imaging_2025, xu_plasmon_2026}.
%The damping behavior could exist even if the deconfined TPT is first-order, as it occupies a finite range of the phase diagram across the phase boundary. Furthermore, we discuss direct experimental signatures such as light transmittance and electric conductance near the deconfined TPT between CFL and FL. 
%In the following, we first summarize our results and describe the physical picture, and then present the detailed calculations. %\ASc{Cite the works of Di Xiao, and of Shankar and Murthy, on FCIs}

% M: traditional picture of CFL is that the resisitivity is given by the composite fermion and static CS terms from the Chern Simons term; near transition flux becomes dynamics; parton pictue is useful to describe this picture. 
% Perhaps first describe fixed points of CFL and FL; and then say that we use Partons to describe the transition

\begin{figure}[!t]
    \centering
    \includegraphics[width = 240pt]{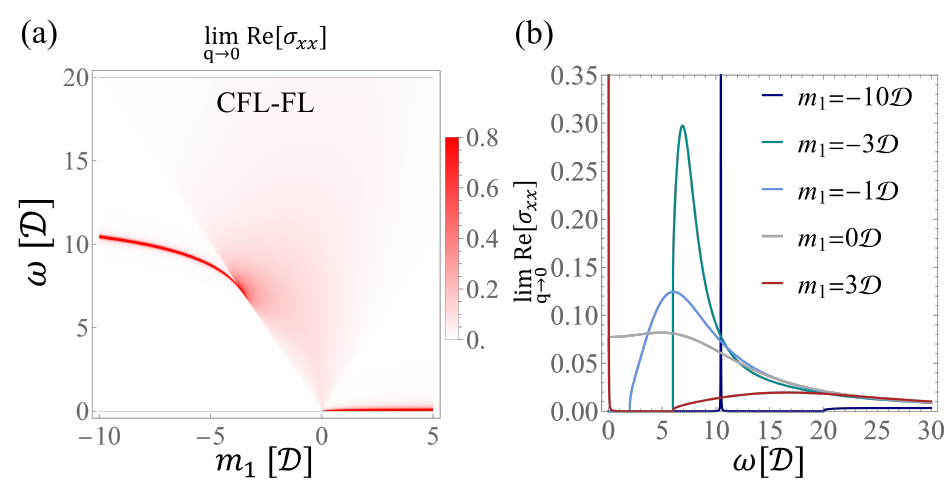}
    \caption{\textit{Damping of the plamon mode near a CFL-FL topological phase transition.}  (a) %\ASc{Why don't we say "The real part of the electronic $\sigma_{xx}$ for $q=0$ as a function of the frequency $\omega$?"} 
    Absolute value of the real part of the conductivity $\sigma_{xx}(\omega)$ in the vicinity of the CFL ($m_1<0$) to FL ($m_1>0$) transition in the long wavelength limit, $q\to 0$. %We set the Drude weight of the $f$ parton Fermi liquid $D=1$ in the simulation. 
    The plasma mode has a long lifetime at large $\abs{m_1}$ but is damped when it enters the particle-hole continuum of the emergent Dirac fermions near the deconfined topological phase transition (the fan is determined by $\omega> 2\abs{m_1}$). (b) Line cuts at different masses $m_1$. %\ASc{Replace $\Pi$ by $\sigma$ in the figures, and $m$ by $\omega_2$.}
    The energies are measured in units of the parton Drude weight $\cal D$.
    }
    \label{plasmon_cfl}
\end{figure}

\textit{Plasmon damping near topological critcality}.---The CFL phase in a Chern band~\cite{dong_composite_2023, Goldman_zerofield_2023, stern_transport_2024, abouelkomsan_compressible_2025} may be described by composite fermion theory. In that theory, the relevant quasiparticle is the Composite Fermion (CF), which is composed of an electron and two flux quanta~\cite{jain_composite_2007,doi:10.1142/11751}. The resistivity of the electrons is given by %the sum of the resistivity of the FL formed by CFs and the resistivity of the flux quanta,
\begin{equation}
    \rho_e=\rho_\text{\tiny CF}+\rho_\text{\tiny CS}
    \label{ILRule}
\end{equation}
where 
\begin{equation}
    \rho_\text{\tiny CF}=\begin{bmatrix}
    -\frac{i \omega}{\cal D} & 0\\
    0 & -\frac{i \omega}{\cal D}
\end{bmatrix} \quad \text{and} \quad \rho_\text{\tiny CS}=\begin{bmatrix}
    0 & -4\pi\\
    4\pi & 0
    \end{bmatrix}
\label{eq:rhocf}
\end{equation} are the resistivity of the FL formed by the CF, and the resistivity generated by the motion of the two flux quanta (resulting from the Chern-Simons (CS) term of the topological field theory), respectively. Here, the components of the matrix refer to the spatial $x$- and $y$-directions, and we set $e=\hbar=c=1$, such that $4\pi$ corresponds to $2h/e^2$. For now, we focus on zero momentum $q=0$, and neglect the Hall conductivity of the composite fermions, arising from the Berry curvature in their band. Furthermore, we assume a perfectly clean system. 
%In the $qv_F\ll \omega$ limit (with $q$ the wave-vector, $\omega$ the frequency, and $v_F$ the Fermi velocity), 
  Then, the composite fermions have a Drude weight ${\cal D}\equiv \lim_{\omega\rightarrow 0}i\omega \left[\rho_\text{\tiny CF}^{-1}\right]_{xx}$.   These simplifications will be relaxed at a later stage of our analysis.
  
The plasmon corresponds to the zero determinant of the electron resistivity tensor $\rho_e$. Combining Eqs.~(\ref{ILRule}) and (\ref{eq:rhocf}), we see that the plasmon is gapped, at $\omega_p=4\pi \cal D$. The gapping of the plasma mode is a remarkable property of the CFL, which comes hand in hand with the vanishing of the electronic Drude weight, ${\cal D}_e=\lim_{\omega\rightarrow 0}i\omega\sigma^e_{{xx}}$, with $\sigma^e\equiv \rho_e^{-1}$ being the electronic conductivity. By contrast, on the FL side, the plasma mode is gapless at $q=0$, and follows a dispersion $\omega\propto q^{1/2}$. 

Our analysis of the transition between CFL and FL builds on the pioneering work of Barkeshli and McGreevy~\cite{barkeshli_continuous_2012}, which was followed up by Song, Zhang, and Senthil~\cite {song_phase_2024}. 
We replace the resistivity matrix of the flux quanta $\rho_\text{\tiny CS}$ by the sum of two resistivity matrices, $\rho_1,\rho_2$. These resistivity matrices describe particles, i.e., partons, that fully fill two Chern bands, with Chern numbers $\mathcal{C}_1,\mathcal{C}_2$. 
Formally, this is achieved by decomposing the electron into three partons, $c = f d_1 d_2$, where $f$ corresponds to the CF and the combined $d_1$ and $d_2$ to the two flux quanta in the CF description. This approach allows us to treat these flux quanta dynamically. The dynamics is captured by three Lagrangians of the partons, which are coupled to one another by two gauge fields ${ a_\mu},{ b_\mu}$, with $\mu=0,x,y$. We use bold-faced ${\bf a,b}$ for the spatial components of $a_\mu,b_\mu$. The $d_1$ partons are coupled to the gauge field ${ a_\mu}$, the $d_2$ partons are coupled to ${ b_\mu}$, and the $f$ partons are coupled to $-a_\mu-b_\mu$. When $a_\mu$ and $b_\mu$ are integrated out, the density and currents of $f$ are enforced to equal those of $d_1$ and $d_2$. The sum of the three Lagrangians is then, 
\begin{equation}
    \mathcal{L}=\mathcal{L}_\text{\tiny CF}[f^\dagger,f,A_\mu-a_\mu-b_\mu]+\mathcal{L}_1[d_1^\dagger,d_1,a_\mu]+\mathcal{L}_2[d_2^\dagger,d_2,b_\mu]
    \label{Lagrangians}
\end{equation}
with 
\begin{equation}
    \mathcal{L}_1=d_1^\dagger(\partial_t-a_0-\mu)d_1-H_1(d_1^\dagger,d_1, {\bf a}).
    \label{Lagrangian1}
\end{equation}
Here, $H_1$ is the Hamiltonian of the $d_1$ partons. Similar expressions hold for $\mathcal{L}_2$ and $\mathcal{L}_\text{\tiny CF}$. 
Crucially, only the $f$ partons couple to the electromagnetic vector potential $A_\mu$. Thus, the charge density is $f^\dagger f$.
The physical electronic resistivity is given by the Ioffe-Larkin rule to be the sum of the resistivities of the three partons (see Eqs.~\eqref{PiRho} and \eqref{IoffeLarkin} in the End Matter for a derivation)
\begin{equation}\rho_e=\rho_\text{\tiny CF}+\rho_1+\rho_2.
\label{ILRrhos}
\end{equation}
For the CFL phase, we have $\mathcal{C}_1=\mathcal{C}_2=1$, such that the resistivity of the partons in the filled Chern band, at low frequency, is
\begin{equation}
   \rho_{\alpha}=\begin{bmatrix}
    i\frac{ \omega}{ \Delta_{\alpha}} & -\frac{2 \pi}{\mathcal{C}_{\alpha}}\\
    \frac{2 \pi}{\mathcal{C}_{\alpha}} & i\frac{\omega}{ \Delta_{\alpha}}
\end{bmatrix}, 
\label{rhoi}
\end{equation} where $\alpha=1,2$ is the parton number, and $\Delta_{\alpha}$ is proportional to the energy gap of the Chern band filled by the partons $\alpha$. In the limit of $\Delta_{\alpha}\rightarrow\infty$, composite fermion theory is reproduced, including the gapped plasma mode and a vanishing electron Drude weight. 

The Fermi liquid phase is obtained for $-\mathcal{C}_1=\mathcal{C}_2=1$, i.e., when the sum of the Chern numbers of the two partons adds to zero. Then, the combined resistivity of the two parton bands is the resistivity matrix of a superfluid, and the total electron resistivity is dominated by that of the composite fermions. As a consequence, the plasma mode is gapless, and when the gaps $\Delta_\alpha$ are large, the electronic Drude weight is the same as that of the composite fermions, ${\cal D}_e = {\cal D}$. 

Within this framework, a transition from the CFL to the FL requires the Chern number of one of the partons, which we choose to be $\mathcal{C}_1$, to change by two. We may envision this to happen in a direct transition or through an intermediate phase with $\mathcal{C}_1=0$. We first focus on the direct transition. 
To analyze the electromagnetic responses of the CFL and the FL transition, we do not integrate out $a_\mu,b_\mu$, but rather integrate out first the fermionic fields $d_1$ and $d_2$. The effective actions for both bands then depend on the gauge fields $a_\mu,b_\mu$. For the $d_1,d_2$ bands to be gapped at $\nu=1/2$, the flux associated with the Chern-Simons magnetic fields ${\bf \nabla}\times {\bf a}$ and ${\bf \nabla}\times{\bf b}$ should be one half of a flux quantum per lattice cell, such that the magnetic unit cell becomes two lattice cells. Then, the effective action for both $d_1$ and $d_2$ is that of a full Chern band with a two-fold degeneracy of the spectrum. %, protected by translational symmetry~\cite{song_phase_2024}. 
In the vicinity of the phase transition, the gap closing and Chern number changing of the partons is, in general, described by massive Dirac fermions with masses $|m_\alpha| \propto \Delta_\alpha$ flipping their sign.
Concretely, the continuous transition from the CFL to the FL phase takes place by a band inversion of two massive Dirac fermions with equal mass $m_1'=m_1''\equiv m_1$. In our conventions, $m_1 < 0$ corresponds to the CFL phase, and $m_1 >0$ to the FL phase. Note that the simultaneous closure of the mass of both fermions is protected by translational symmetry~\cite{song_phase_2024}, i.e., by the magnetic unit cell being twice the lattice unit cell. 
Throughout the transition, we assume the gap $\Delta_2$ to remain large, and the phase transition is only driven by $\Delta_1$.
At the one-loop level, we express the low-energy effective Lagrangian obtained after integrating out a parton in a gapped Chern insulator in terms of a current-current correlation function, which is a sum of a CS term and a Maxwell term. %For the $d_1$ parton described by two degenerate massive Dirac fermions near the TPT, the one-loop effective Lagrangian is
%$\mathcal{L}_1=\frac{1}{2}a_\mu\Pi_{\mu\nu} a_\nu$, with $\Pi_{\mu\nu}=\Pi_{\mu\nu}^\text{\tiny MW} +\Pi_{\mu\nu}^\text{\tiny CS}$. 
The explicit expression is presented in the End Matter; see Eqs.~\eqref{LagrangianParton1} and \eqref{Pis}. %The resistivity is then obtained from $\rho = T \Pi^{-1} T$, where $T=\text{diag}(\sqrt{iq^2/\omega},\sqrt{i\omega})$.

On the FL side, since $\mathcal{C}_1+\mathcal{C}_2=0$, the plasma mode is gapless. Furthermore, near the transition, the electronic Drude weight, ${\cal D}_e^{-1}={\cal D}^{-1}+\Delta_1^{-1}$, rises continuously from zero when tuning further into the FL phase. On the CFL side, the proximity of the transition affects the plasma mode. For frequencies $\omega>\Delta_1$, the resistivity also has a real, dissipative component, originating from inter-parton-band transitions. 
Thus, close to the critical point, where $\Delta_1$ is small, the plasma mode becomes damped and acquires a width. 
In Fig.~\ref{plasmon_cfl} we present the real part of the conductivity at $q=0$ as a function of $\omega$, calculated from Eq.~(\ref{ILRrhos}) for different values of $\Delta_1$, parameterized by $m_1$. The sharply-defined gapped plasmon is observed at $m_1<-\pi\mathcal{D}$. Its spectral weight is smeared when $-\pi\mathcal{D}<m_1<0$, where the plasmon mode enters the particle-hole continuum of the $d_1$ parton. This damping rate increases as the critical point is approached, and thus could be distinguished from the impurity scattering, which contributes a nearly constant damping rate. The plasmon becomes gapless for positive $m_1$, implying a non-zero Drude weight, which grows linearly with $m_1$ close to the critical point (see Eq.~\eqref{drudeweight_FL} in the End matter). 
%Our calculation neglects the decay of a $q=0$ plasmon to two particle-hole excitations of $\pm q$ within the continuum of particle-hole excitations of the $f$-partons. This decay would result in a non-zero plasmon width, even deep in the CFL phase. 
We remark that our parton theory captures only intra-band processes and neglects electronic transitions to higher Chern bands. As a result, a portion of the spectral weight in the physical response will be transferred to inter-band transitions at higher energies. In our theory, the spectral weight resides entirely in the shown intra-band excitations. 
The finite spectral weight for intra-band excitations at $q=0$ is allowed because we consider a Chern band~\cite{Wu_moire_2016, paul_shining_2025, shen_magnetorotons_2025, kousa_theory_2025}, where continuous translation symmetry is broken, which should be contrasted with a Landau level with Galilean invariance. 

% Next, we analyze the critical point ($m_1=0$). Here, the electronic polarization operator is
% %$\Pi_{1}=\begin{bmatrix}
% %    \frac{i q^2}{16\omega} & 0\\
%  %   0& \frac{i \omega^2}{16\omega}
%  %   \end{bmatrix}$ in the Coulomb gauge, and consequently
% \begin{equation}
% \Pi^e=\frac{1}{(\frac{\omega}{\mathcal{D}}+8i)^2-4\pi^2}\begin{bmatrix}
%     -(\frac{1}{\mathcal{D}}+\frac{8i}{\omega})q^2 & -2\pi iq\\
%     2\pi iq & -(\frac{\omega^2}{\mathcal{D}}+8i\omega)
%     \end{bmatrix}.
% \end{equation}
% The real part of the longitudinal conductivity  at small $\omega$ is then $\frac{2}{16+\pi^2}$, 
% %\begin{equation}
%  %   \lim_{\omega\rightarrow 0}\Im[\Pi_e^{00}]=\frac{1}{16+\pi^2}
% %\end{equation}
% which agrees with the linecuts shown in Fig.~\ref{plasmon_cfl}(b).  

At the critical point ($m_1=0$) we find that the conductivity satisfies an ``f-sum rule", $\lim_{q\rightarrow 0}\int_{-\infty}^{\infty}d\omega \Re[\sigma^e_{xx}(\omega)]=\pi \mathcal{D}$, even though we use a relativistic Dirac fermion polarization tensor in composing the electron response function; see Eqs.~\eqref{Pi_at_m=0} and \eqref{sigma_at_m=0} in the End Matter for details. However, this should not be understood as the familiar f-sum rule where the integral gives $\pi \frac{n}{m_e}$ with $m_e$ the bare mass of electrons, because the $\mathcal{D}$ we used for the $f$-parton Fermi liquid generally depends on the interaction energy scale. 
%\mkch{(except deep in the FL regime)}\XYc{you mean deep in FL side?}. 
%\XYt{Notice that there is only a single parameter $\mathcal{D}$ in expression ~\eqref{Pi_at_m=0} for the optical conductivity, a direct comparison between the experimentally measured conductivity and Eq.~\eqref{Pi_at_m=0} at low energy would be useful to extract the Drude weight at the critical point.}

%We note that we use free Dirac fermions as the critical theory, which do not faithfully reflect the critical behavior. 
We use free Dirac fermions as the critical theory, which provides a good leading-order approximation.
%, but cannot quantitatively reflect the critical behavior.
Since we focus on the qualitative behavior at low energies and near the bandgap, we argue that the general features still persist in a more accurate critical theory. Remarkably, the mechanism for plasmon damping is insensitive to the precise values of the critical exponents. Moreover, we expect this damping to happen even when the transition is weakly first-order, since the plasmon damping arises in a finite range around the critical point.

{\it Other transport signatures}.---Two other physical phenomena whose behavior across the transition we explore are the chirality-selective transmission of light (``chiral mirror''), which occurs at a frequency $\omega_0$ for which $\det \sigma_e(\omega_0)=0$, and the longitudinal electronic conductivity at $q\ne 0$, whose linear dependence on $q$ is an identifying feature of composite Fermi liquid physics. The first requires us to incorporate the Berry curvature of the composite fermion band into $\rho_\text{\tiny CF}$, while the second requires the incorporation of the wave vector $q$ into the expressions we derived above. 
\begin{figure}
    \centering
    \includegraphics[width = 240pt]{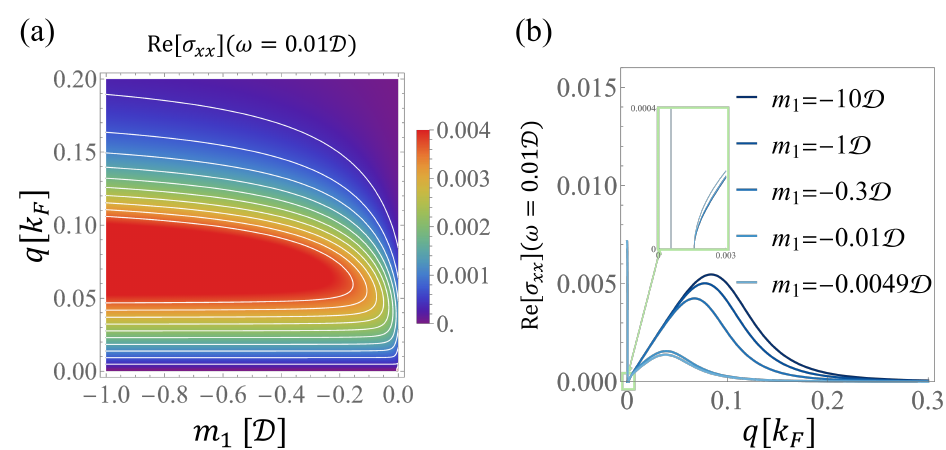}
    \caption{\textit{Dissipative part of conductivity at finite momentum.} (a) Electron conductivity at $\omega=0.01 {\cal D}$ in CFL ($m_1<0$) phase near the phase transition at different wave vectors.  %When the contours are nearly equal distant, the conductivity $\Re\sigma_{xx}$ is linear in $q$. 
    The region with a linear in $q$ conductivity (approximately equal distant contours) is bounded by $|m_1|$ when $|m_1|$ is small. (b) Line cuts at different $m_1$. The inset shows the behavior at small $q$ indicated by the green box. When the mass of the Dirac fermion $m_1$ goes below $\omega/2$, we also get a real conductivity at small $q$ due to the dissipation events from the particle-hole continuum of the almost gapless Dirac fermion parton.
    }
    \label{cond_cfl}
\end{figure}

When the CF band has a finite Berry curvature, which integrates to a non-zero Hall conductance $\sigma_{xy}^\text{\tiny CF}$ in a half-filled band, the CF resistivity becomes 
\begin{equation}
    \rho_\text{\tiny CF}=\begin{bmatrix}
    -i\frac{ \cal D}{ \omega} & \sigma_{xy}^\text{\tiny CF}\\
    -\sigma_{xy}^\text{\tiny CF} & -i\frac{\cal D}{ \omega}
\end{bmatrix}^{-1}
\end{equation}
 and its determinant diverges for $\omega_0={\cal D}/{\sigma_{xy}^\text{\tiny CF}}$. 
The reflection and transmission amplitudes $E_r$ and $E_t$ of an electromagnetic wave of amplitude $E_0$ incident perpendicular to a (2D) system are closely related to the conductivity matrix $\sigma_e$
\begin{eqnarray}
    E_r&=-&\frac{1}{1+{\tilde\alpha\sigma_e}}{\tilde\alpha\sigma_e} E_0 \\
    E_t&=&\frac{1}{1+{\tilde\alpha\sigma_e}}E_0
    \label{reflection}
\end{eqnarray}
where $\tilde\alpha$ is the fine structure constant~\cite{tse_magneto-optical_2011, liu_anomalous_2020, stern_transport_2024}. 
%When $\det \sigma_e(\omega_0)=0$ for some frequency $\omega_0$, we obtain perfect transmission for one direction of circular polarization, d
Consequently, for $\det \sigma_e(\omega_0)=0$, we obtain perfect transmission for one direction of circular polarization (determined by the eigenvector of the zero eigenvalue of $\sigma_e$).
The frequency at which that happens, $\omega_0$, is continuous across the transition. This is because $\rho_{1,2}$, being the resistivities of a full Chern band,  are not singular at any frequency. Hence, the chirality-selective transmission originates from the vanishing of the determinant of the conductivity of the $f$ partons only. 
 
 %Since , the determinant of the electronic resistivity diverges at $\omega_0$ on both sides of the transition, and consequently, $\det \sigma=0$ at that frequency on both sides of the transition. %\ASc{Can $\det\sigma_e$ be zero at new frequencies due to the parton's resistivities?}\XYc{No, it could only from CF, assuming other partons are Dirac/have infinity gap}

To analyze the electronic conductivity at $q\ne 0$, we set $\sigma_{xy}^{\text{\tiny CF}}=0$ again. 
%\ASc{Can anything interesting happen if the CFs have $\sigma_{xy}$?}\XYc{For small q, the diagonal components win}. 
A characteristic property of the CFL is that its conductivity is linear in wave vector~\cite{halperin_theory_1993,willett_enhanced_1993}, $\Re\sigma^e_{xx}\propto q$, when $16\pi^2 (q/k_F)^3<\omega/{\cal D}< 4\pi q/k_F$ with $k_F$ the Fermi momentum. 
%\frac{4\pi }{k_F m^*}q^3<\omega< \frac{k_F}{m^*} q$ with $k_F$ the Fermi momentum and $m^*$ the effective mass of the $f$ partons{\color{red}?}, and related to the Drude weight $\cal D$ through $\mathcal{D}=\frac{k_F^2}{4\pi m^*}$.
% and $n$ the density of the composite fermion FL. 
% This $q$ dependent conductivity is a consequence of the CF resistivity at $\frac{k_F}{nm^*}q^3<\omega< \frac{k_F}{m^*} q$ being $\rho_{\text{\tiny CF}}^{-1}\approx\begin{bmatrix}
%     -i\frac{m^*}{2\pi}\frac{\omega}{q^2} & 0\\
%     0& \frac{k_F}{2\pi q}
% \end{bmatrix}$. 
Near the transition, the gap gets small, and $\rho_\text{\tiny CS}$ changes dramatically when a transition of a single parton from a  full to an empty parton band becomes possible at the wave vector $q$ and frequency $\omega$. 
%becomes comparable to the gap of the flux quanta. 
Therefore, the breakdown of the linear in $q$ behavior of $\Re \sigma^e_{xx}$ allows us to infer when that happens. This behavior is seen in Fig.~(\ref{cond_cfl}); see End Matter for details of the calculation. 
%Eqs.~(\ref{finiteq_Pi_f}-\ref{complicated_sigma}) in the

%$q$ is comparable to the gap scale of the flux quanta, see Fig.~\ref{cond_cfl}.\ASc{I'm not sure what you mean - $q$ and the gap do not have the same units}\XYc{we set the velocity of Dirac fermion to 1}

%{\it The calculation}.---After describing the behavior of several transport quantities throughout a continuous CFL-FL transition, we now turn to present the outline of the calculation. 

{\it Compressibility and magnetic susceptibility}.---Thermodynamic quantities demonstrate interesting behavior near the critical point. To extract the compressibility and magnetic susceptibility, we need to consider the zero-frequency value of the electron electromagnetic response, which is related to the electromagnetic response of the partons by the inverse sum rule (Eq.~\eqref{IoffeLarkin} and Eq.~\eqref{ILRule}). %This relation has the same origin as the resistivity (inverse conductivity) sum rule in Eq.~\eqref{ILRule}, but f
For thermodynamic quantities, we first set $\omega=0$ and then evaluate the $ q\rightarrow 0$ limit (see End Matter for a detailed derivation). In this limit, one finds that the compressibility vanishes as one approaches the critical point, 
\begin{equation}
    \frac{\partial n}{\partial \mu}\approx\frac{3\abs{m_1}}{\pi}
\end{equation} with 
%$n$ the electron density, $\mu$ the chemical potential, $m^*$ the effective mass of the $f$ parton and 
$m_1$ the mass  of the $d_1$ parton that drives the phase transition. The incompressible critical point was pointed out by Barkeshli and McGreevy~\cite{barkeshli_continuous_2012}, and it comes from the critical dynamics of the $d_1$ parton with a singular magnetic susceptibility. 

Furthermore, the magnetic susceptibility $\chi$ is immune to the singular magnetic susceptibility of the $d_1$ parton. We find $\chi$ to be approximately the Landau diamagnetic susceptibility of the $f$ parton near the critical point,
\begin{equation}
    \chi=\frac{\partial M}{\partial B}\approx-\frac{1}{12\pi m^*}
\end{equation} with $M$ the electron orbital magnetization, $B$ the external magnetic field and $m^*$ the effective mass of $f$ parton. Within our approximation, $\chi_\text{\tiny FL}=-\frac{1}{12\pi m^*}$ deep in the Fermi liquid side, while deep in the CFL phase $\chi_\text{\tiny CFL}=-\frac{1}{20\pi m^*}$.  This is an interesting behavior: while the precise values of the susceptibility may change with interaction, the crucial observation is that $\chi$ remains finite at the transition. The half-filled Chern band at the critical point has thus vanishing compressibility yet a non-zero magnetic susceptibility. 
%as the vanishing(tiny) Drude weight on the CFL(FL close to the critical point) implies a vanishing(tiny) quasiparticle residue, which determines the transport properties, while the magnetic susceptibility comes from all the electrons below the chemical potential and does not decrease as the Drude weight vanishes.

{\it Indirect transitions}.---So far, we considered a CFL-FL transition in which one of the bands of $d_1$ or $d_2$ goes through a transition between $\mathcal{C}=+1$ and $\mathcal{C}=-1$ states. In principle, this transition may be indirect and go through a $\mathcal{C}=0$ phase. Such a transition requires breaking of translational symmetry, such that the lattice unit cell is doubled~\cite{barkeshli_anyon_2010, song_phase_2024}. A full band with $\mathcal{C}=0$ is a trivial insulator, for which the resistivity is infinite at zero temperature. Such a band would then dominate the Ioffe-Larkin rule Eq.~\eqref{ILRrhos}, and lead to a perfectly insulating state at $\nu=1/2$. The scenario of an indirect transition, however, may also lead to a different outcome. Since Chern bands generally lack electron-hole symmetry around $\nu=1/2$, there is a difference between a parton state of the type we consider when composed of electrons and when composed of holes. In the case of electrons, the electronic resistivity we calculate is the system's electronic resistivity. In the case of holes, the resistivity we calculate should be inverted to a conductivity, which then adds to the conductivity of the full Chern band. Hence, when the resistivity is infinite, the holes have a vanishing conductivity, and the total conductivity becomes that of the full Chern band. Consequently, the intermediate phase, which we call insulator*, has vanishing longitudinal charge conductivity, but its Hall conductivity may be finite.

\begin{figure}[!t]
    \centering
    \includegraphics[width = 240pt]{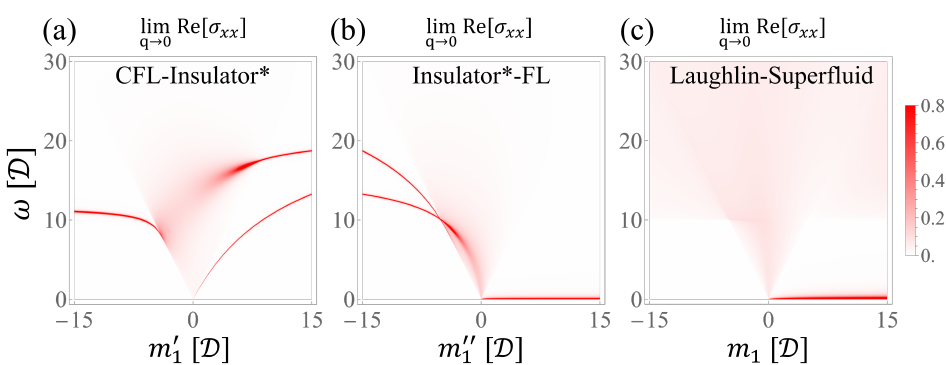}
    \caption{\textit{Plasmon damping for different deconfined TPTs.} %Absolute value of the imaginary part of electron density-density response function divided by $q^2$ in the long wavelength limit in the vicinity of different deconfined TPTs. 
    Near the deconfined TPTs, the plasma mode gets damped when entering the particle-hole continuum of the Dirac fermions (fan defined by $\omega> 2\abs{m_1}$), similar to the CFL-FL transition. (a) CFL-insulator* transition, where one of the Dirac fermions ($m_1'$) flips the sign of the mass, while the other retains a negative mass, $m_1''<0$. (b) Insulator*-FL transition, where $m_1'$ remains positive while $m_1''$ flips its sign. (c) Laughlin $\nu=1/2$ state ($m_1<0$) to superfluid ($m_1>0$) transition.  %assume the $d_2$ parton to be a Dirac fermion with mass $-5\mathcal{D}$, and 
    There is no sharp plasma mode for the Laughlin  $\nu=1/2$ state in a Chern band. %In (a) and (b), we assumed the $d_2$ parton to have an infinite gap. 
    }
     \label{plasmon_l}
\end{figure}

%If the transition of the fermion $d_1$ from $\mathcal{C}=1$ to $\mathcal{C}=-1$ is indirect, and goes through a trivial $\mathcal{C}=0$ insulator state the transition from a CFL to a FL is modified into a transition of CFL-insulator*-FL. 
While the insulator* has an infinite charge resistivity at zero temperature due to the infinite resistivity of the $d_1$ parton, it still carries heat due to the gapless excitations of the $f$-partons~\cite{lee_gauge_1992,barkeshli_continuous_2012}. 
%\footnote{It is an electrically insulating and incompressible state while being thermally conductive~\cite{lee_gauge_1992,barkeshli_continuous_2012}.}-FL)\ASc{How can a state carry heat and be incompressible?}
%, where the transition happens when one of the Dirac fermions flips its mass sign. 
Since in the insulator* phase lattice translational symmetry is broken, the masses of the two Dirac fermions of $d_1$, $m_1'$, and $m_1''$ are generally different.
%Denoting the mass of the two Dirac fermions of $d_1$ by $m_1'$ and $m_1''$ respectively, 
This has consequences on the evolution of the plasma mode in the long wavelength limit across the CFL-insulator*-FL transition; see Fig.~\ref{plasmon_l}(a) and (b).
In Fig.~\ref{plasmon_l}(a), we show the plasmon behavior near the critical point ($m_1'=0$, $m_1''<0$) of the CFL-insulator* TPT, where we change the mass of one of the Dirac fermions ($m_1'$) while keeping the other ($m_1''=-15 \mathcal{D}$) unchanged and keeping $m_2$ large throughout.
The plasmon behavior close to the insulator*-FL critical point ($m_1'>0$, $m_1''=0$) is shown in Fig.~\ref{plasmon_l}(b), where we change $m_1''$ while keeping $m_1'=15\mathcal{D}$ unchanged. In the insulator* phase, we observe two distinct plasma modes, which appear due to the distinct masses $m_1' \neq m_1''$, related to broken translations.
While the details vary compared to the simple CFL-FL transition, we find that the plasmon damping is a rather general and robust feature of TPTs.

\textit{Plasma modes in bosonic deconfined TPTs}.---The CFL-FL is only one of several TPTs that may take place within a partially filled Chern band. We now turn to discuss the plasma modes when the Chern band is half-filled with bosons, and the transition is from the Laughlin $\nu=1/2$ to a superfluid~\cite{barkeshli_continuous_2014}. In this case, the deconfined TPT theory proposes to fractionalize the boson into two fermionic partons, $b^\dagger=d_1^\dagger d_2^\dagger$. When both $d_1$ and $d_2$ are in insulating states with $\mathcal{C}=1$, the boson is in the Laughlin state. By contrast, when $d_1$ ($d_2$) is in an insulating state with $\mathcal{C}=-1$ ($\mathcal{C}=1$), the corresponding bosonic state is a superfluid. Therefore, the Laughlin-superfluid transition results from a continuous TPT of $d_1$ from $\mathcal{C}=1$ to $\mathcal{C}=-1$. The picture is analogous to that of the transition between CFL and FL, as the two transitions are argued to share the same critical theory~\cite{barkeshli_continuous_2012}.
%when we fractionalize the electron into a fermionic parton $f$ and a bosonic parton $b$, $c^\dagger=f^\dagger b^\dagger$, 
%and the Laughlin-superfluid transition is argued to share the same critical theory as CFL-FL~\cite{barkeshli_continuous_2012}. 

The plasma mode of this transition deserves discussion. %Unlike the case of free electrons in a magnetic field, where Galilean invariance guarantees the existence of a sharp plasma mode (Kohn's mode), no such guarantee exists for a Chern band.  
%For the Laughlin $\nu=1/2$ state of bosons in a half-filled Chern band to have a sharp plasma mode, this mode should reside in the gap between the filled and empty partons bands. 
Deep in the Laughlin state side, when the bands of the two partons are identical, the electronic resistivity is just twice the resistivity matrix of a filled Chern band. A full Chern band has a continuum of interband particle-hole excitations, but no sharp collective plasma mode. This is because the determinant of the resistivity does not vanish at frequencies within the gap, but rather at the threshold of the particle-hole continuum. Near the transition, the gaps of the two bands become very different, and additional continuous particle-hole excitations of the parton with the smaller energy gap become possible. Therefore, no sharp collective mode arises. 
Between these limiting cases, a sharp plasma mode may occur in principle, but it is not guaranteed to occur.
We show the conductivity in Fig.~\ref{plasmon_l}(c) using the Dirac cone response functions, leading to a broad continuum instead of a sharp plasma mode for all values of the gaps of the two partons. In the calculations, we assumed the $d_2$ parton to be a Dirac fermion with mass $m_2=-5\mathcal{D}$.
%We find the damping behavior is very similar to the CFL-FL transition: the sharp gapped plasmon mode far from the phase transition gets damped by the critical particle-hole continuum of Dirac fermions.

%\XYc{The thermodynamic quantities in the insulator* phase are worth mentioning. Unlike conventional band insulators, where the electric polarizability and diamagnetic susceptibility both scale inversely with the gap, in the insulator* phase, these two quantities can behave very differently. More precisely, close to the transition between insulator* and FL/CFL, the electric polarizability is given by $\alpha=\frac{1}{\Delta_1}$, with $\Delta_1$ the gap scale of the $d_1$ parton, while the diamagnetic susceptibility is given by $\chi=-\frac{1}{12\pi m^* +\Delta_1}$. This shows that as one approaches the critical point from the insulator* phase, $\Delta_1$ decreases to zero, the electric polarizability diverges, while the diamagnetic susceptibility saturates.}

% \ASc{Let's discuss what to plot}
% In the Laughlin state, too, the frequency of the plasmon mode depends on the gaps of $d_1$ and $d_2$ in parton theory. Here $d_2$ is a Dirac fermion with a gap scale $\Delta_2$,\ASc{Let's talk again about the footnote}\footnote{We still assume an infinitesimal length scale for the CS term to be valid at all length scales for simplicity.} which is set by introducing a Maxwell term in the polarization tensor of $d_2$, $\Pi_2^\text{\tiny MW}=\frac{1}{\Delta}\left(-\eta_{\mu\nu}q_\alpha q^\alpha+q_\mu q_\nu\right)$. We plot the plasmon mode in Fig.~\ref{plasmon_l}(a) with $\Delta=100\cal D$. We find the damping behavior is very similar to the CFL-FL transition: the sharp gapped plasmon mode far from the phase transition gets damped by the critical particle-hole continuum of Dirac fermions.

{\it Summary \& Outlook}.---We discussed the behavior of distinct electromagnetic-response observables across a continuous topological phase transition between a CFL and an FL in the half-filled Chern band, as well as several related transitions. Recent experiments have found indications for a rich landscape of unconventional phases~\cite{cai_signatures_2023,park_observation_2023,xu_observation_2023,zeng_thermodynamic_2023, xu_signatures_2025, lu_fractional_2024,han_large_2024,han_signatures_2025,lu_extended_2025, liu_fractional_2025} and their electromagnetic response can be directly characterized with THz spectroscopy~\cite{xu_electronic_2024,chen_direct_2025, chen_terahertz_2025, pierce_imaging_2025, xu_plasmon_2026}.
Our discussion was based on two premises -- the addition of resistivities of the three partons, which is a defining feature of a parton system, and the description of the partons $d_1$ and $d_2$ in terms of massive Dirac fermions. The second premise is approximate, but its central features, in particular the behavior of the response functions for $\omega$ close to zero and close to the energy gap are a consequence of general considerations. %Although the continuous evolution we found for the plasmon, the Drude weight, the chiral mirroring, and the $q$-dependent conductivity can become discontinuous should the transition be of first-order, we expect the general features we discuss to prevail, as they arise in an extended region around the critical point. 

For future work, it will be interesting to numerically study the transition and quantitatively predict the spectral weights of the conductivity in concrete settings. In addition, the formalism we develop allows us to predict the electromagnetic response away from a filling of $\nu=1/2$, where a rich landscape of topological phases and exotic phase transitions may arise. In particular, superconductivity has been predicted to appear in the vicinity of topological phase transitions~\cite {divic_anyon_2025, pichler_microscopic_2025, kuhlenkamp_robust_2025, chen_topological_2025}, whose electromagnetic response would be interesting to study.

\begin{acknowledgments}
{\it Acknowledgments}.---We thank Long Ju, Zach Hadjri, Clemens Kuhlenkamp, and Di Xiao for fruitful discussions. X.Y. thanks Wei Ku, Tingxin Li, Guorui Chen, and Xiaoxue Liu for helpful discussions and Xin Liu at TDLI for hospitality, where this work was initiated. We acknowledge support from the Israeli Science Foundation, the Israeli Ministry of Science Technology and Space, the Minerva Stiftung, the
DFG (CRC/Transregio 183, EI 519/7-1), the Israel Science Foundation ISF (Grant No 1914/24), ISF Quantum Science and Technology (2074/19), from the Deutsche Forschungsgemeinschaft (DFG, German Research Foundation) under Germany’s Excellence Strategy--EXC--2111--390814868, TRR 360 – 492547816 and DFG grants No. KN1254/1-2, KN1254/2-1, the European Union (grant agreement No. 101169765), as well as the Munich Quantum Valley (MQV), which is supported by the Bavarian state government with funds from the Hightech Agenda Bayern Plus. 
\end{acknowledgments}

{\it{Data availability.---}}All data is contained in this manuscript.

\section{End Matter}
\subsection{Current-current response function}
To calculate the conductivity and resistivity of the $d_1$ parton described by two degenerate massive Dirac fermions, we first compute the effective Lagrangian near the TPT. On a one-loop level, we find 
\begin{equation}
    \mathcal{L}_1[a_\mu]=\frac{1}{2}a_\mu\Pi_{\mu\nu} a_\nu, \label{LagrangianParton1}
\end{equation}
%To calculate the conductivity and resistivity, we write the 
%For the $d_1$ parton described by two degenerate massive Dirac fermions near the TPT, the one-loop effective Lagrangian is
%$\mathcal{L}_1=\frac{1}{2}a_\mu\Pi_{\mu\nu} a_\nu$, with $\Pi_{\mu\nu}=\Pi_{\mu\nu}^\text{\tiny MW} +\Pi_{\mu\nu}^\text{\tiny CS}$. 
where the current-current response function of massive Dirac fermions $\Pi_{\mu \nu} = \Pi_{\mu\nu}^\text{\tiny MW} + \Pi_{\mu\nu}^\text{\tiny CS}$ is composed of a Maxwell and Chern-Simons term, given by
\begin{widetext}
\begin{equation}
\begin{aligned}
    \Pi_{\mu\nu}^\text{\tiny MW}&=\frac{1}{4\pi}\left(-\eta_{\mu\nu}q_\alpha q^\alpha+q_\mu q_\nu\right) \frac{1}{q_E^3}\left(2q_E\abs{m}-(4m^2-q_E^2) \arctan(\frac{q_E}{2\abs{m}})\right)\\
    \Pi_{\mu\nu}^\text{\tiny CS}&=-\frac{im}{2\pi}\epsilon_{\mu\lambda\nu}q^\lambda \frac{2}{q_E}\arctan(\frac{q_E}{2\abs{m}})
    \label{Pis}
\end{aligned}
%\label{PisDC}
\end{equation}
\end{widetext}
with the Dirac velocity set to one for simplicity, $\eta_{\mu\nu}=\text{diag}(1,-1,-1)$, $q_E=\sqrt{-\omega^2+\bf{q}^2}$, and $m$ the mass of Dirac fermion. We choose the UV regularizations for the two Dirac fermions so that there is no extra CS term. Similar expressions hold for the second parton, with $a_\mu$ being replaced by $b_\mu$ and the mass changed to the $d_2$ parton band gap accordingly. 

In the Coulomb gauge, where ${\bf q}\cdot{\bf a}={\bf q}\cdot{\bf b}=0$, and for $\bf q||\hat{x}$, the correlation function reduces to 
\begin{equation}
    \Pi_{\mu\nu}=\begin{bmatrix}
    \Pi_{00} & \Pi_{0y}\\
    \Pi_{y0} & \Pi_{yy}
    \label{twobytwoPi}
    \end{bmatrix}
\end{equation}
    %\ASc{Xinlei, $\Pi$ is 2*2 or 3*3?}\XYc{It's 3x3, and in Coulomb gauge it reduces to 2x2, should we just put the result under coulomb gauge here?}\ASc{Yes, I think that would be better, given that the resistivity is 2*2} 

From the continuity equation, it follows that the polarization operator $\Pi$ is related to the resistivity matrices $\rho$  by  
\begin{equation}\rho=\begin{bmatrix}
    \sqrt{\frac{iq^2}{\omega}} & 0\\
    0 & \sqrt{i\omega}
\end{bmatrix} \Pi^{-1} \begin{bmatrix}
    \sqrt{\frac{iq^2}{\omega}} & 0\\
    0 & \sqrt{i\omega}
\end{bmatrix}.
\label{PiRho}
\end{equation}
Note that the components of the resistivity matrices referred to the spatial $x$- and $y$-directions, while the components of $\Pi$ refer to temporal and spatial $y$ indices as in Eq.~\eqref{twobytwoPi}.
% with $T=\begin{bmatrix}
%     \sqrt{\frac{\omega}{iq^2}} & 0\\
%     0 & \sqrt{\frac{-i}{\omega}}
% \end{bmatrix}.$    

The electronic response function $\Pi_e$, i.e., the response function to the electromagnetic field $A_\mu$, is obtained by intergrating out the fields $a_\mu,b_\mu$
\begin{equation}
    \Pi_e^{-1}=\Pi_f^{-1}+\Pi_{1}^{-1}+\Pi_{2}^{-1}.
    \label{IoffeLarkin}
\end{equation}
with $\Pi_\alpha$ being the response function of the $d_\alpha$ parton, and $\Pi_f$ being the response function of the $f$-parton. 
Together with Eq.~(\ref{PiRho}), this relation leads to the Ioffe-Larkin rule Eq.~\eqref{ILRrhos}.

In the low-frequency and long-wavelength limit, $m^2\gg q^2,\omega^2$, the polarization operator $\Pi_{\mu\nu}$ is dominated by the CS term, $\frac{\mathcal{C}}{4\pi}ada$ with $\mathcal{C}=-1$ ($\mathcal{C}=1$) for $m>0$ ($m<0$), corresponding to a Hall conductance of $\pm 1$. When that is the case, the $a_0$ terms in Eq.~\eqref{Lagrangian1}  enforce the constraints ${\bf \nabla}\times{\bf a}=C_1d_1^\dagger d_1$ and ${\bf \nabla}\times{\bf b}=C_2d_2^\dagger d_2$. Then, when $C_1=C_2=1$, the $f$-partons experience a magnetic field corresponding to two flux quanta per electron, making the $f$-partons composite fermions. Since $\nu=1/2$, this field corresponds to one flux quantum per lattice unit cell, such that the magnetic unit cell and the lattice unit cell coincide and the $f$ parton remains a gapless FL state. The electron system then forms a CFL. 
%In contrast, when $C_1=-C_2$, the $f$-partons do not experience any magnetic field, and form an electronic Fermi liquid. \XYc{let's discuss this}

The diagonal term of $\Pi$ is translated by Eq.~(\ref{PiRho}) to an imaginary diagonal resistivity that is proportional to $\omega$ at low frequencies, and approaches a constant as $\omega$ approaches the energy gap, where it also acquires a real, dissipative part. The linear dependence of the diagonal conductivity on $i\omega$ holds for small $\omega$ for all gapped systems. The details of its evolution for frequencies well within the gap depend on the details of the band structure, which are not captured by the effective Dirac cone model.     

For $m_1=0$, at the phase transition, the parton $d_1$ Lagrangian includes only the Maxwell term $\frac{1}{8}\left(-\eta_{\mu\nu}q_\alpha q^\alpha+q_\mu q_\nu\right)\frac{1}{q_E}$.
In the long-wavelength limit $q\rightarrow 0$ we set $\omega\rightarrow \omega+i\delta$ with $\delta$ infinitesimal to get the retarded function $[\Pi_1]_{00}=\frac{1}{8}\frac{\abs{\vec{q}}^2}{-i \omega}$. For two gapless Dirac fermions, this gives the well-known constant optical conductivity of the $d_1$ partons via $\sigma_{xx}(\omega)=\frac{\omega}{i\abs{\vec{q}}^2}\Pi_{00}=\frac{1}{8}$ and the electronic current-current response reduces to 

 % Next, we analyze the critical point ($m_1=0$). Here, the electronic polarization operator is
% $\Pi_{1}=\begin{bmatrix}
%    \frac{i q^2}{16\omega} & 0\\
%    0& \frac{i \omega^2}{16\omega}
%    \end{bmatrix}$ in the Coulomb gauge, and consequently
\begin{equation}
\Pi^e=\frac{1}{(\frac{\omega}{\mathcal{D}}+8i)^2-4\pi^2}\begin{bmatrix}
    -(\frac{1}{\mathcal{D}}+\frac{8i}{\omega})q^2 & -2\pi iq\\
    2\pi iq & -(\frac{\omega^2}{\mathcal{D}}+8i\omega)
    \end{bmatrix}. \label{Pi_at_m=0}
\end{equation}
The real part of the longitudinal conductivity  at small $\omega$ is then
\begin{equation}
   \lim_{\omega\rightarrow 0}\Re[\sigma^e_{xx}]=\frac{2}{16+\pi^2}, \label{sigma_at_m=0}
\end{equation}
which agrees with the linecuts shown in Fig.~\ref{plasmon_cfl}(b).  In the small frequency limit, the real part is $\Re[\sigma^e_{xx}(\omega)]=\frac{2}{16+\pi^2}+\frac{-16+3\pi^2}{2(16+\pi^2)^3}\frac{\omega^2}{\mathcal{D}^2}$ and the imaginary part increase linearly with frequency $\Im[\sigma^e_{xx}(\omega)]=\frac{16-\pi^2}{4(16+\pi^2)^2}\frac{i\omega}{\mathcal{D}}$.

Eqs.~(\ref{Pis}) and (\ref{IoffeLarkin}) are what is needed for the analysis of the transport properties we discussed in the main text, and their behavior when tuning through the phase transition. 
The plasmon and the Drude weight are directly obtained from these expressions by setting  $q\rightarrow0$, and using Eq.~(\ref{PiRho}) to obtain the resistivity. The plasmon is a sharp mode when the diagonal element of the resistivity is purely imaginary, and the Hall element is real. This requires the frequency of the mode to be within the gap of the two parton bands. Close to the transition, this condition ceases to hold, and the plasmon gets damped.

\subsection{Drude weight and $q$-dependent conductivity}
We now calculate the Drude weight and the $q$-dependent conductivity close to the CFL-FL transition. 

 %Using our formalism, we now determine the conductivity near the phase transition. 
 A characteristic quantity of the FL is the electronic Drude weight, ${\cal D}_e=\lim_{\omega\rightarrow 0}i\omega\sigma^e_{xx}(q=0,\omega)$, with $\sigma^e\equiv \rho_e^{-1}$ being the electronic conductivity, in the absence of impurity scattering. We start from the FL ($m_1>0$) side, in the low frequency limit, $m_1\gg \omega\gg q$, we have $\Pi_{1}=\frac{1}{12\pi m_1}\begin{bmatrix}
    q^2 & 0\\
    0& \omega^2
    \end{bmatrix}+\frac{1}{2\pi}\begin{bmatrix}
    0 & -iq\\
    iq& 0
    \end{bmatrix}.$
This leads to an electron polarization
\begin{equation}
    \Pi_{e}=-\frac{1}{\frac{1}{\mathcal{D}}+\frac{\pi}{3m_1}}\begin{bmatrix}
    \frac{q^2}{\omega^2} & 0\\
    0 & 1 
    \end{bmatrix}
    \label{drudeweight_FL}
\end{equation}
which describes an FL with Drude weight $\frac{1}{\frac{1}{D}+\frac{\pi}{3m_1}}$. We find the Drude weight saturates at $\mathcal D$ for large $m_1$ and decreases continuously to zero as $m_1\rightarrow 0$. The linear decrease of Drude weight at small $m_1$ is a result of a free Dirac fermion, which could possibly be modified by the critical theory 

A unique property of the CFL phase is the wave vector-dependent conductivity $\Re\sigma^e_{xx}\propto q$ for ${16\pi^2 }(q/k_F)^3<\omega/\mathcal{D}<4 \pi q/k_F$ with $k_F$ the Fermi momentum. 
To extract the linear in $q$ conductivity in the CFL ($m_1<0$) phase, we need the polarization tensors for ${16\pi^2 }(q/k_F)^3<\omega/\mathcal{D}<4 \pi q/k_F$. In this limit, the $f$ Fermi liquid contributes 
% \begin{equation}
%     \Pi_{f}\approx \begin{bmatrix}
%     \frac{m^*}{2\pi} & 0\\
%     0& \frac{2in\omega}{k_F q}
%     \end{bmatrix}.
% \end{equation}
$\Pi_{f}\approx \begin{bmatrix}
    \frac{k_F^2}{8\pi^2 \mathcal{D}} & 0\\
    0& \frac{i\omega}{2\pi}\frac{k_F}{q}
    \end{bmatrix}. $
%\label{finiteq_Pi_f}
When $q\ll \abs{m_1}$, the $d_1$ Dirac fermions contribute $\Pi_{1}=\begin{bmatrix}
    \frac{q^2}{12\pi \abs{m_1}} & \frac{iq}{2\pi}\\
    -\frac{iq}{2\pi} & -\frac{q^2}{12\pi \abs{m_1}}
    \end{bmatrix}$
and we arrive at the linear in $q$ conductivity $\sigma^e_{xx}(q,\omega)=\frac{q}{2\pi k_F}$, while for $q\gg \abs{m_1}$ we find
$\Pi_{1}=\begin{bmatrix}
    \frac{q}{16} & -i\frac{m_1}{4}\\
    i\frac{m_1}{4} & -\frac{q}{16}
    \end{bmatrix}$,
which gives us $\sigma^e_{xx}(q,\omega)=\frac{i\omega\left(\pi\left(16 m_1^{2}-q^{2}\right)-8 i\,\omega\,k_{F}\right)}
{\left(16 q + 8\pi^2\mathcal{D}\frac{q^2}{k_F^2}\right)(\pi q^2 + 8i k_F\omega) + 128\pi^2 \mathcal{D}m_1^2 \frac{q^2}{k_F^2}} \label{complicated_sigma}$
%\begin{widetext} 
% \begin{equation}
%     \sigma_{xx}(q,\omega)=\frac{i\,m^*\,\omega\left(\pi\left(-16 m^{2}+q^{2}\right)+8 i\,\omega\,k_{F}\right)}
% {-2\pi q^2\left(-16 m^{2}\pi+q\left(8\,m^*+\pi q\right)\right)-16 i\left(8\,m^*+\pi q\right)\omega\,k_{F}q}
% \end{equation}
%\begin{equation}
%    \sigma_{xx}(q,\omega)=\frac{i\omega\left(\pi\left(16 m_1^{2}-q^{2}\right)-8 i\,\omega\,k_{F}\right)}
%{\left(16 q + 8\pi^2\mathcal{D}\frac{q^2}{k_F^2}\right)(\pi q^2 + 8i k_F\omega) + 128\pi^2 \mathcal{D}m_1^2 %\frac{q^2}{k_F^2}} \label{complicated_sigma}
%\end{equation}
%\end{widetext}
with a complicated $(q,\omega)$ dependence. Thus, the linear in $q$ conductivity $\sigma^e_{xx}$ terminates at around scale $q_c\sim \abs{m_1}$ as shown in Fig.~\ref{cond_cfl}. %In the simulation, we set the $f$ parton Fermi liquid to have an effective mass $m^*=1$ and Fermi wavevector $k_F=\sqrt{4\pi}$ so that the Drude weight of the $f$ parton Fermi liquid is $D=\frac{k_F^2}{4\pi}=1$. 
Starting from a large Dirac fermion mass $m_1$, when ${4 \pi {\cal D}} q/k_F >\omega$, we get a real part of conductivity from the particle-hole continuum of the $f$ parton Fermi liquid and when $16\pi^2 {\cal D} (q/k_F)^3\ll \omega$, $\Re\sigma^e_{xx}\propto q$. When the gap of the Dirac fermion, $2m$, gets below $\omega$, we get additional real conductivity at small $q$ from the dissipation channels of the Dirac fermions. 
    
    %We also note that when $\omega\gg \frac{q^2}{k_F}$, $\Re[\sigma_{xx}]\propto q$, which may be observed near the critical region where $\abs{m}\sim 0$. However, the detailed behavior depends strongly on the microscopic details.
\subsection{Compressibility and Magnetic Susceptibility}
Thermodynamic quantities can also be extracted from Eq.~\eqref{IoffeLarkin}, but now in the $\omega=0$, $q\rightarrow 0$ limit. The polarization of the $f$ parton Fermi liquid is $\Pi_{f}=\begin{bmatrix}
    \frac{m^*}{2\pi} & 0\\
    0 & -\frac{q^2}{12\pi m^*}
    \end{bmatrix}$. Together with the polarization tensor of $d_1$ with mass $m_1$, $\Pi_{1}=\begin{bmatrix}
    \frac{q^2}{12\pi \abs{m_1}} & \pm\frac{iq}{2\pi}\\
    \mp\frac{iq}{2\pi} & -\frac{q^2}{12\pi \abs{m_1}}
    \end{bmatrix}$ with the upper(lower) sign corresponding to the CFL(FL) phase, and $d_2$, which is assumed to contribute $\Pi_{2}=\frac{1}{2\pi}\begin{bmatrix}
    0 & iq\\
    -iq & 0
    \end{bmatrix}$ for simplicity, we find that the electron polarization tensor is given by
    \begin{equation}
    \begin{aligned}
        \Pi_e=&\frac{q^2}{\left(\frac{2\pi}{m^*}+\frac{\pi}{3\abs{m_1}}\right)12\pi m^*+\left(2\pi\pm2\pi\right)^2}\\
        &\times\begin{bmatrix}
    \frac{12\pi m^*}{q^2} & i\frac{2\pi\pm2\pi}{q}\\
    -i\frac{2\pi\pm2\pi}{q} & -\left(\frac{2\pi}{m^*}+\frac{\pi}{3\abs{m_1}}\right)
    \end{bmatrix}.
    \end{aligned}
    \end{equation}
From this, we extract the compressibility (magnetic susceptibility) through $\Pi_e^{00}$ ($\Pi_e^{yy}/q^2$). 

For the insulator* phase, the only difference is that $d_1$ parton does not have a Hall response and $\Pi_{1}=\begin{bmatrix}
    \frac{q^2}{\Delta_1} & 0\\
    0 & -\frac{q^2}{\Delta_1}
    \end{bmatrix}$ with $\Delta_1$ the gap scale of the $d_1$ parton and $\Delta_1\approx 12\pi \min(\abs{m_1'},\abs{m_1''})$. This, in turn, gives an incompressible state with
    \begin{equation}
        \Pi_e=\begin{bmatrix}
    \frac{q^2}{\Delta_1} & 0\\
    0 & -\frac{q^2}{12\pi m^* +\Delta_1}
    \end{bmatrix}
    \end{equation}
which has the same form as a band insulator, but we note that the relation between electric polarizability and magnetic susceptibility is non-trivial. As one approaches the phase transition, $\Delta_1\rightarrow 0$, indicating a diverging electric polarizability, while the magnetic susceptibility saturates due to the $f$ parton Fermi surface, in contrast to a conventional band insulator.

\subsection{Plasma modes in the bosonic Laughlin and insulator* phases}
In this section, we discuss the behavior of plasma modes in the bosonic Laughlin $\nu=1/2$ state and insulator* phase. For the bosonic $\nu=1/2$ Laughlin state,  we find that there is no sharp plasma mode, just as in the case of a fermionic fully-filled Chern band.  This is because here the plasmon frequency is higher than the gap of the $d_2$ parton, and therefore it is damped even far from the phase transition. 

For the insulator* phase, where the two Dirac cones of the $d_1$ parton have two different masses, we find two plasma modes. The easiest way to understand this finding is from looking at the polarization tensor when the masses of the Dirac fermions are opposite $m_1'=-m_1''$. In this case, the polarization tensor for $d_1$ parton behaves as $\Pi_{1}=\frac{1}{12\pi \abs{m_1'}}\begin{bmatrix}
    q^2 & 0\\
    0& \omega^2
    \end{bmatrix}$, together with the contribution $\Pi_{f}=-\mathcal{D}\begin{bmatrix}
    \frac{q^2}{\omega^2} & 0\\
    0 & 1
    \end{bmatrix}$ from $f$ parton and $\Pi_{2}=\frac{1}{2\pi}\begin{bmatrix}
    0 & iq\\
    -iq & 0
    \end{bmatrix}$ from $d_2$ parton. We note that in this limit, the signs of the polarization tensor of the $d_1$ parton and the $d_2$ parton are different. Physically, this arises from the different responses of metals and insulators to an external electric field. For a metal, the quasiparticles have a mass, and therefore the current response to the external electric field is delayed, which corresponds to a positive imaginary part of the conductivity. On the other hand, for an insulator, a low-frequency electric field induces a polarization, and the current is the time derivative of this polarization, which corresponds to a negative imaginary conductivity. The plasmon frequencies are then determined by $\det\Pi_e^{-1}=0$, which further reduces to $\abs{\frac{12\pi \abs{m_1'}}{\omega}-\frac{\omega}{\mathcal{D}}}=2\pi$; an equation that has two positive solutions $\omega_{p_{1,2}}$. These considerations also hold more generally for different masses of the Dirac fermions $m_1'\neq m_1''$.

\bibliography{references-2}

@article{wu_bipartite_2025,
	title = {Bipartite {Fluctuations} of {Critical} {Fermi} {Surfaces}},
	volume = {15},
	issn = {2160-3308},
	url = {https://link.aps.org/doi/10.1103/dflw-rksw},
	doi = {10.1103/dflw-rksw},
	language = {en},
	number = {3},
	urldate = {2026-02-08},
	journal = {Physical Review X},
	author = {Wu, Xiao-Chuan},
	month = aug,
	year = {2025},
	pages = {031035},
}

@article{wang_nearly_2011,
	title = {Nearly flat band with {Chern} number \${C}=2\$ on the dice lattice},
	volume = {84},
	url = {https://link.aps.org/doi/10.1103/PhysRevB.84.241103},
	doi = {10.1103/PhysRevB.84.241103},
	number = {24},
	urldate = {2026-02-03},
	journal = {Physical Review B},
	publisher = {American Physical Society},
	author = {Wang, Fa and Ran, Ying},
	month = dec,
	year = {2011},
	pages = {241103},
}

@article{liu_anomalous_2020,
	title = {Anomalous {Hall} effect, magneto-optical properties, and nonlinear optical properties of twisted graphene systems},
	volume = {6},
	copyright = {2020 The Author(s)},
	issn = {2057-3960},
	url = {https://www.nature.com/articles/s41524-020-0299-4},
	doi = {10.1038/s41524-020-0299-4},
	language = {en},
	number = {1},
	urldate = {2026-01-04},
	journal = {npj Computational Materials},
	author = {Liu, Jianpeng and Dai, Xi},
	month = may,
	year = {2020},
	note = {Publisher: Nature Publishing Group},
	pages = {57},
}

@article{abouelkomsan_quantum_2023,
	title = {Quantum metric induced phases in {Moiré} materials},
	volume = {5},
	issn = {2643-1564},
	url = {https://link.aps.org/doi/10.1103/PhysRevResearch.5.L012015},
	doi = {10.1103/PhysRevResearch.5.L012015},
	language = {en},
	number = {1},
	urldate = {2025-11-27},
	journal = {Physical Review Research},
	author = {Abouelkomsan, Ahmed and Yang, Kang and Bergholtz, Emil J.},
	month = feb,
	year = {2023},
	pages = {L012015},
}

@article{wen_continuous_2000,
	title = {Continuous {Topological} {Phase} {Transitions} between {Clean} {Quantum} {Hall} {States}},
	volume = {84},
	copyright = {http://link.aps.org/licenses/aps-default-license},
	issn = {0031-9007, 1079-7114},
	url = {https://link.aps.org/doi/10.1103/PhysRevLett.84.3950},
	doi = {10.1103/PhysRevLett.84.3950},
	language = {en},
	number = {17},
	urldate = {2025-02-05},
	journal = {Physical Review Letters},
	author = {Wen, Xiao-Gang},
	month = apr,
	year = {2000},
	pages = {3950--3953},
}

@article{qi_generic_2011,
	title = {Generic {Wave}-{Function} {Description} of {Fractional} {Quantum} {Anomalous} {Hall} {States} and {Fractional} {Topological} {Insulators}},
	volume = {107},
	url = {https://link.aps.org/doi/10.1103/PhysRevLett.107.126803},
	doi = {10.1103/PhysRevLett.107.126803},
	number = {12},
	urldate = {2025-10-12},
	journal = {Physical Review Letters},
	author = {Qi, Xiao-Liang},
	month = sep,
	year = {2011},
	note = {Publisher: American Physical Society},
	pages = {126803},
}

@article{barkeshli_topological_2012,
	title = {Topological {Nematic} {States} and {Non}-{Abelian} {Lattice} {Dislocations}},
	volume = {2},
	url = {https://link.aps.org/doi/10.1103/PhysRevX.2.031013},
	doi = {10.1103/PhysRevX.2.031013},
	number = {3},
	urldate = {2025-01-27},
	journal = {Physical Review X},
	author = {Barkeshli, Maissam and Qi, Xiao-Liang},
	month = aug,
	year = {2012},
	pages = {031013},
}

@article{moller_fractional_2015,
	title = {Fractional {Chern} {Insulators} in {Harper}-{Hofstadter} {Bands} with {Higher} {Chern} {Number}},
	volume = {115},
	url = {https://link.aps.org/doi/10.1103/PhysRevLett.115.126401},
	doi = {10.1103/PhysRevLett.115.126401},
	number = {12},
	urldate = {2025-10-18},
	journal = {Physical Review Letters},
	author = {Möller, Gunnar and Cooper, Nigel R.},
	month = sep,
	year = {2015},
	note = {Publisher: American Physical Society},
	pages = {126401},
}

@article{wu_bloch_2013,
	title = {Bloch {Model} {Wave} {Functions} and {Pseudopotentials} for {All} {Fractional} {Chern} {Insulators}},
	volume = {110},
	url = {https://link.aps.org/doi/10.1103/PhysRevLett.110.106802},
	doi = {10.1103/PhysRevLett.110.106802},
	number = {10},
	urldate = {2025-10-18},
	journal = {Physical Review Letters},
	author = {Wu, Yang-Le and Regnault, N. and Bernevig, B. Andrei},
	month = mar,
	year = {2013},
	note = {Publisher: American Physical Society},
	pages = {106802},
}

@article{kapit_exact_2010,
	title = {Exact {Parent} {Hamiltonian} for the {Quantum} {Hall} {States} in a {Lattice}},
	volume = {105},
	url = {https://link.aps.org/doi/10.1103/PhysRevLett.105.215303},
	doi = {10.1103/PhysRevLett.105.215303},
	number = {21},
	urldate = {2025-10-12},
	journal = {Physical Review Letters},
	author = {Kapit, Eliot and Mueller, Erich},
	month = nov,
	year = {2010},
	note = {Publisher: American Physical Society},
	pages = {215303},
}

@article{murthy_hamiltonian_2012,
	title = {Hamiltonian theory of fractionally filled {Chern} bands},
	volume = {86},
	url = {https://link.aps.org/doi/10.1103/PhysRevB.86.195146},
	doi = {10.1103/PhysRevB.86.195146},
	number = {19},
	urldate = {2025-10-14},
	journal = {Physical Review B},
	author = {Murthy, Ganpathy and Shankar, R.},
	month = nov,
	year = {2012},
	note = {Publisher: American Physical Society},
	pages = {195146},
}

@article{neupert_fractional_2011,
	title = {Fractional {Quantum} {Hall} {States} at {Zero} {Magnetic} {Field}},
	volume = {106},
	url = {https://link.aps.org/doi/10.1103/PhysRevLett.106.236804},
	doi = {10.1103/PhysRevLett.106.236804},
	number = {23},
	urldate = {2025-10-12},
	journal = {Physical Review Letters},
	author = {Neupert, Titus and Santos, Luiz and Chamon, Claudio and Mudry, Christopher},
	month = jun,
	year = {2011},
	note = {Publisher: American Physical Society},
	pages = {236804},
}

@article{tang_high-temperature_2011,
	title = {High-{Temperature} {Fractional} {Quantum} {Hall} {States}},
	volume = {106},
	url = {https://link.aps.org/doi/10.1103/PhysRevLett.106.236802},
	doi = {10.1103/PhysRevLett.106.236802},
	number = {23},
	urldate = {2025-10-12},
	journal = {Physical Review Letters},
	author = {Tang, Evelyn and Mei, Jia-Wei and Wen, Xiao-Gang},
	month = jun,
	year = {2011},
	note = {Publisher: American Physical Society},
	pages = {236802},
}

@article{regnault_fractional_2011,
	title = {Fractional {Chern} {Insulator}},
	volume = {1},
	url = {https://link.aps.org/doi/10.1103/PhysRevX.1.021014},
	doi = {10.1103/PhysRevX.1.021014},
	number = {2},
	urldate = {2025-10-12},
	journal = {Physical Review X},
	author = {Regnault, N. and Bernevig, B. Andrei},
	month = dec,
	year = {2011},
	note = {Publisher: American Physical Society},
	pages = {021014},
}

@article{hu_hyperdeterminants_2024,
	title = {Hyperdeterminants and composite fermion states in fractional {Chern} insulators},
	volume = {109},
	url = {https://link.aps.org/doi/10.1103/PhysRevB.109.245125},
	doi = {10.1103/PhysRevB.109.245125},
	number = {24},
	urldate = {2025-10-14},
	journal = {Physical Review B},
	author = {Hu, Xiaodong and Xiao, Di and Ran, Ying},
	month = jun,
	year = {2024},
	note = {Publisher: American Physical Society},
	pages = {245125},
}

@article{dong_composite_2023,
	title = {Composite {Fermi} {Liquid} at {Zero} {Magnetic} {Field} in {Twisted} {MoTe} 2},
	volume = {131},
	issn = {0031-9007, 1079-7114},
	url = {https://link.aps.org/doi/10.1103/PhysRevLett.131.136502},
	doi = {10.1103/PhysRevLett.131.136502},
	language = {en},
	number = {13},
	urldate = {2025-10-14},
	journal = {Physical Review Letters},
	author = {Dong, Junkai and Wang, Jie and Ledwith, Patrick J. and Vishwanath, Ashvin and Parker, Daniel E.},
	month = sep,
	year = {2023},
	pages = {136502},
}

@misc{shen_magnetorotons_2025,
	title = {Magnetorotons in {Moiré} {Fractional} {Chern} {Insulators}},
	url = {http://arxiv.org/abs/2412.01211},
	doi = {10.48550/arXiv.2412.01211},
	urldate = {2025-10-12},
	publisher = {arXiv},
	author = {Shen, Xiaoyang and Wang, Chonghao and Hu, Xiaodong and Guo, Ruiping and Yao, Hong and Wang, Chong and Duan, Wenhui and Xu, Yong},
	month = feb,
	year = {2025},
	note = {arXiv:2412.01211 [cond-mat]},
}

@misc{paul_shining_2025,
	title = {Shining light on collective modes in moiré fractional {Chern} insulators},
	url = {http://arxiv.org/abs/2502.17569},
	doi = {10.48550/arXiv.2502.17569},
	urldate = {2025-10-12},
	publisher = {arXiv},
	author = {Paul, Nisarga and Abouelkomsan, Ahmed and Reddy, Aidan and Fu, Liang},
	month = feb,
	year = {2025},
	note = {arXiv:2502.17569 [cond-mat]},
}

@article{abouelkomsan_compressible_2025,
	title = {Compressible {Quantum} {Liquid} with {Vanishing} {Drude} {Weight}},
	volume = {134},
	url = {https://link.aps.org/doi/10.1103/PhysRevLett.134.176501},
	doi = {10.1103/PhysRevLett.134.176501},
	number = {17},
	urldate = {2025-09-16},
	journal = {Physical Review Letters},
	author = {Abouelkomsan, Ahmed and Paul, Nisarga and Stern, Ady and Fu, Liang},
	month = apr,
	year = {2025},
	note = {Publisher: American Physical Society},
	pages = {176501},
}

@article{tse_magneto-optical_2011,
	title = {Magneto-optical {Faraday} and {Kerr} effects in topological insulator films and in other layered quantized {Hall} systems},
	volume = {84},
	url = {https://link.aps.org/doi/10.1103/PhysRevB.84.205327},
	doi = {10.1103/PhysRevB.84.205327},
	number = {20},
	urldate = {2025-10-09},
	journal = {Physical Review B},
	author = {Tse, Wang-Kong and MacDonald, A. H.},
	month = nov,
	year = {2011},
	note = {Publisher: American Physical Society},
	pages = {205327},
}

@article{lee_gauge_1992,
	title = {Gauge theory of the normal state of high- {T} c superconductors},
	volume = {46},
	copyright = {http://link.aps.org/licenses/aps-default-license},
	issn = {0163-1829, 1095-3795},
	url = {https://link.aps.org/doi/10.1103/PhysRevB.46.5621},
	doi = {10.1103/PhysRevB.46.5621},
	language = {en},
	number = {9},
	urldate = {2024-05-17},
	journal = {Physical Review B},
	author = {Lee, Patrick A. and Nagaosa, Naoto},
	month = sep,
	year = {1992},
	pages = {5621--5639},
}

@book{doi:10.1142/11751,
	title = {Fractional quantum hall effects},
	url = {https://www.worldscientific.com/doi/abs/10.1142/11751},
	publisher = {WORLD SCIENTIFIC},
	author = {Halperin, Bertrand I and Jain, Jainendra K},
	year = {2020},
	doi = {10.1142/11751},
	note = {tex.eprint: https://www.worldscientific.com/doi/pdf/10.1142/11751},
}

@book{jain_composite_2007,
	address = {Cambridge},
	title = {Composite {Fermions}},
	isbn = {978-0-521-86232-5},
	url = {https://www.cambridge.org/core/books/composite-fermions/AB22E09A3F9C4E98F91E1BB447AF5778},
	urldate = {2025-10-08},
	publisher = {Cambridge University Press},
	author = {Jain, Jainendra K.},
	year = {2007},
	doi = {10.1017/CBO9780511607561},
}

@misc{yahuizhang_continuous_2025,
	title = {Continuous transition from {Fermi} liquid to {A} fractional {Chern} insulator},
	url = {http://arxiv.org/abs/2507.22130},
	doi = {10.48550/arXiv.2507.22130},
	urldate = {2025-08-06},
	publisher = {arXiv},
	author = {Zhang, Ya-Hui},
	month = jul,
	year = {2025},
	note = {arXiv:2507.22130 [cond-mat]},
}

@article{willett_enhanced_1993,
	title = {Enhanced finite-wave-vector conductivity at multiple even-denominator filling factors in two-dimensional electron systems},
	volume = {47},
	copyright = {http://link.aps.org/licenses/aps-default-license},
	issn = {0163-1829, 1095-3795},
	url = {https://link.aps.org/doi/10.1103/PhysRevB.47.7344},
	doi = {10.1103/PhysRevB.47.7344},
	language = {en},
	number = {12},
	urldate = {2025-10-08},
	journal = {Physical Review B},
	author = {Willett, R. L. and Ruel, R. R. and Paalanen, M. A. and West, K. W. and Pfeiffer, L. N.},
	month = mar,
	year = {1993},
	pages = {7344--7347},
}

@misc{chen_terahertz_2025,
	title = {Terahertz electrodynamics in a zero-field {Wigner} crystal},
	url = {http://arxiv.org/abs/2509.10624},
	doi = {10.48550/arXiv.2509.10624},
	urldate = {2025-09-22},
	publisher = {arXiv},
	author = {Chen, Su-Di and Qi, Ruishi and Kim, Ha-Leem and Feng, Qixin and Xia, Ruichen and Abeysinghe, Dishan and Xie, Jingxu and Taniguchi, Takashi and Watanabe, Kenji and Lee, Dung-Hai and Wang, Feng},
	month = sep,
	year = {2025},
	note = {arXiv:2509.10624 [cond-mat]},
}

@article{stern_transport_2024,
	title = {Transport {Properties} of a {Half}-{Filled} {Chern} {Band} at the {Electron} and {Composite} {Fermion} {Phases}},
	volume = {133},
	issn = {0031-9007, 1079-7114},
	url = {https://link.aps.org/doi/10.1103/PhysRevLett.133.246602},
	doi = {10.1103/PhysRevLett.133.246602},
	language = {en},
	number = {24},
	urldate = {2025-09-16},
	journal = {Physical Review Letters},
	author = {Stern, Ady and Fu, Liang},
	month = dec,
	year = {2024},
	pages = {246602},
}

@misc{zhang_continuous_2025,
	title = {Continuous topological phase transition between Z$_2$ topologically ordered phases},
	url = {http://arxiv.org/abs/2508.08376},
	doi = {10.48550/arXiv.2508.08376},
	urldate = {2025-08-17},
	publisher = {arXiv},
	author = {Zhang, Qi and Xu, Wen-Tao},
	month = aug,
	year = {2025},
	note = {arXiv:2508.08376 [cond-mat]},
}

@misc{zhou_chern-simons-matter_2025,
	title = {Chern-{Simons}-matter conformal field theory on fuzzy sphere: {Confinement} transition of {Kalmeyer}-{Laughlin} chiral spin liquid},
	shorttitle = {Chern-{Simons}-matter conformal field theory on fuzzy sphere},
	url = {http://arxiv.org/abs/2507.19580},
	doi = {10.48550/arXiv.2507.19580},
	urldate = {2025-08-06},
	publisher = {arXiv},
	author = {Zhou, Zheng and Wang, Chong and He, Yin-Chen},
	month = jul,
	year = {2025},
	note = {arXiv:2507.19580 [cond-mat]},
}

@misc{lotric_paired_2025,
	title = {Paired {Parton} {Trial} {States} for the {Superfluid}-{Fractional} {Chern} {Insulator} {Transition}},
	url = {http://arxiv.org/abs/2504.20139},
	doi = {10.48550/arXiv.2504.20139},
	urldate = {2025-08-06},
	publisher = {arXiv},
	author = {Lotrič, Tevž and Simon, Steven H.},
	month = apr,
	year = {2025},
	note = {arXiv:2504.20139 [cond-mat]},
}

@misc{wang_emergent_2025,
	title = {Emergent {QED}$_3$ at the bosonic {Laughlin} state to superfluid transition},
	url = {http://arxiv.org/abs/2507.07611},
	doi = {10.48550/arXiv.2507.07611},
	urldate = {2025-08-06},
	publisher = {arXiv},
	author = {Wang, Taige and Song, Xue-Yang and Zaletel, Michael P. and Senthil, T.},
	month = jul,
	year = {2025},
	note = {arXiv:2507.07611 [cond-mat]},
}

@misc{xu_signatures_2025,
	title = {Signatures of unconventional superconductivity near reentrant and fractional quantum anomalous {Hall} insulators},
	url = {http://arxiv.org/abs/2504.06972},
	doi = {10.48550/arXiv.2504.06972},
	urldate = {2025-08-06},
	publisher = {arXiv},
	author = {Xu, Fan and Sun, Zheng and Li, Jiayi and Zheng, Ce and Xu, Cheng and Gao, Jingjing and Jia, Tongtong and Watanabe, Kenji and Taniguchi, Takashi and Tong, Bingbing and Lu, Li and Jia, Jinfeng and Shi, Zhiwen and Jiang, Shengwei and Zhang, Yuanbo and Zhang, Yang and Lei, Shiming and Liu, Xiaoxue and Li, Tingxin},
	month = apr,
	year = {2025},
	note = {arXiv:2504.06972 [cond-mat]},
}

@misc{han_exotic_2024,
	title = {Exotic phase transitions in spin ladders with discrete symmetries that emulate spin-1/2 bosons in two dimensions},
	url = {http://arxiv.org/abs/2412.17911},
	doi = {10.48550/arXiv.2412.17911},
	urldate = {2025-08-05},
	publisher = {arXiv},
	author = {Han, Bo and Mross, David F.},
	month = dec,
	year = {2024},
	note = {arXiv:2412.17911 [cond-mat]},
}

@article{han_large_2024,
	title = {Large quantum anomalous {Hall} effect in spin-orbit proximitized rhombohedral graphene},
	volume = {384},
	url = {https://www.science.org/doi/full/10.1126/science.adk9749},
	doi = {10.1126/science.adk9749},
	number = {6696},
	urldate = {2025-08-03},
	journal = {Science},
	author = {Han, Tonghang and Lu, Zhengguang and Yao, Yuxuan and Yang, Jixiang and Seo, Junseok and Yoon, Chiho and Watanabe, Kenji and Taniguchi, Takashi and Fu, Liang and Zhang, Fan and Ju, Long},
	month = may,
	year = {2024},
	pages = {647--651},
}

@article{zeng_thermodynamic_2023,
	title = {Thermodynamic evidence of fractional {Chern} insulator in moiré {MoTe2}},
	volume = {622},
	copyright = {2023 The Author(s), under exclusive licence to Springer Nature Limited},
	issn = {1476-4687},
	url = {https://www.nature.com/articles/s41586-023-06452-3},
	doi = {10.1038/s41586-023-06452-3},
	language = {en},
	number = {7981},
	urldate = {2025-08-03},
	journal = {Nature},
	author = {Zeng, Yihang and Xia, Zhengchao and Kang, Kaifei and Zhu, Jiacheng and Knüppel, Patrick and Vaswani, Chirag and Watanabe, Kenji and Taniguchi, Takashi and Mak, Kin Fai and Shan, Jie},
	month = oct,
	year = {2023},
	pages = {69--73},
}

@article{park_observation_2023,
	title = {Observation of fractionally quantized anomalous {Hall} effect},
	volume = {622},
	copyright = {2023 The Author(s), under exclusive licence to Springer Nature Limited},
	issn = {1476-4687},
	url = {https://www.nature.com/articles/s41586-023-06536-0},
	doi = {10.1038/s41586-023-06536-0},
	language = {en},
	number = {7981},
	urldate = {2025-08-03},
	journal = {Nature},
	author = {Park, Heonjoon and Cai, Jiaqi and Anderson, Eric and Zhang, Yinong and Zhu, Jiayi and Liu, Xiaoyu and Wang, Chong and Holtzmann, William and Hu, Chaowei and Liu, Zhaoyu and Taniguchi, Takashi and Watanabe, Kenji and Chu, Jiun-Haw and Cao, Ting and Fu, Liang and Yao, Wang and Chang, Cui-Zu and Cobden, David and Xiao, Di and Xu, Xiaodong},
	month = oct,
	year = {2023},
	pages = {74--79},
}

@article{lu_fractional_2024,
	title = {Fractional quantum anomalous {Hall} effect in multilayer graphene},
	volume = {626},
	copyright = {2024 The Author(s), under exclusive licence to Springer Nature Limited},
	issn = {1476-4687},
	url = {https://www.nature.com/articles/s41586-023-07010-7},
	doi = {10.1038/s41586-023-07010-7},
	language = {en},
	number = {8000},
	urldate = {2025-08-03},
	journal = {Nature},
	author = {Lu, Zhengguang and Han, Tonghang and Yao, Yuxuan and Reddy, Aidan P. and Yang, Jixiang and Seo, Junseok and Watanabe, Kenji and Taniguchi, Takashi and Fu, Liang and Ju, Long},
	month = feb,
	year = {2024},
	pages = {759--764},
}

@article{han_signatures_2025,
	title = {Signatures of chiral superconductivity in rhombohedral graphene},
	volume = {643},
	copyright = {2025 The Author(s), under exclusive licence to Springer Nature Limited},
	issn = {1476-4687},
	url = {https://www.nature.com/articles/s41586-025-09169-7},
	doi = {10.1038/s41586-025-09169-7},
	language = {en},
	number = {8072},
	urldate = {2025-08-03},
	journal = {Nature},
	author = {Han, Tonghang and Lu, Zhengguang and Hadjri, Zach and Shi, Lihan and Wu, Zhenghan and Xu, Wei and Yao, Yuxuan and Cotten, Armel A. and Sharifi Sedeh, Omid and Weldeyesus, Henok and Yang, Jixiang and Seo, Junseok and Ye, Shenyong and Zhou, Muyang and Liu, Haoyang and Shi, Gang and Hua, Zhenqi and Watanabe, Kenji and Taniguchi, Takashi and Xiong, Peng and Zumbühl, Dominik M. and Fu, Liang and Ju, Long},
	month = jul,
	year = {2025},
	pages = {654--661},
}

@article{cai_signatures_2023,
	title = {Signatures of {Fractional} {Quantum} {Anomalous} {Hall} {States} in {Twisted} {MoTe2} {Bilayer}},
	volume = {622},
	issn = {0028-0836, 1476-4687},
	url = {http://arxiv.org/abs/2304.08470},
	doi = {10.1038/s41586-023-06289-w},
	number = {7981},
	urldate = {2025-07-28},
	journal = {Nature},
	author = {Cai, Jiaqi and Anderson, Eric and Wang, Chong and Zhang, Xiaowei and Liu, Xiaoyu and Holtzmann, William and Zhang, Yinong and Fan, Fengren and Taniguchi, Takashi and Watanabe, Kenji and Ran, Ying and Cao, Ting and Fu, Liang and Xiao, Di and Yao, Wang and Xu, Xiaodong},
	month = oct,
	year = {2023},
	pages = {63--68},
}

@article{xu_observation_2023,
	title = {Observation of {Integer} and {Fractional} {Quantum} {Anomalous} {Hall} {Effects} in {Twisted} {Bilayer} {MoTe} 2},
	volume = {13},
	issn = {2160-3308},
	url = {https://link.aps.org/doi/10.1103/PhysRevX.13.031037},
	doi = {10.1103/PhysRevX.13.031037},
	language = {en},
	number = {3},
	urldate = {2025-08-03},
	journal = {Physical Review X},
	author = {Xu, Fan and Sun, Zheng and Jia, Tongtong and Liu, Chang and Xu, Cheng and Li, Chushan and Gu, Yu and Watanabe, Kenji and Taniguchi, Takashi and Tong, Bingbing and Jia, Jinfeng and Shi, Zhiwen and Jiang, Shengwei and Zhang, Yang and Liu, Xiaoxue and Li, Tingxin},
	month = sep,
	year = {2023},
	pages = {031037},
}

@article{barkeshli_continuous_2015,
	title = {Continuous {Preparation} of a {Fractional} {Chern} {Insulator}},
	volume = {115},
	url = {https://link.aps.org/doi/10.1103/PhysRevLett.115.026802},
	doi = {10.1103/PhysRevLett.115.026802},
	number = {2},
	urldate = {2025-07-15},
	journal = {Physical Review Letters},
	author = {Barkeshli, M. and Yao, N. Y. and Laumann, C. R.},
	month = jul,
	year = {2015},
	pages = {026802},
}

@article{lu_continuous_2025,
	title = {Continuous {Transition} between {Bosonic} {Fractional} {Chern} {Insulator} and {Superfluid}},
	volume = {134},
	url = {https://link.aps.org/doi/10.1103/PhysRevLett.134.076601},
	doi = {10.1103/PhysRevLett.134.076601},
	number = {7},
	urldate = {2025-07-14},
	journal = {Physical Review Letters},
	author = {Lu, Hongyu and Wu, Han-Qing and Chen, Bin-Bin and Meng, Zi Yang},
	month = feb,
	year = {2025},
	pages = {076601},
}

@article{pichler_microscopic_2025,
  title = {Microscopic Mechanism of Anyon Superconductivity Emerging from Fractional {{Chern}} Insulators},
  author = {Pichler, Fabian and Kuhlenkamp, Clemens and Knap, Michael and Vishwanath, Ashvin},
  journal = {Newton},
  publisher = {Elsevier},
  doi = {10.1016/j.newton.2025.100340},
  urldate = {2026-01-22}
}

@article{song_phase_2024,
	title = {Phase transitions out of quantum {Hall} states in moir{\textbackslash}'e materials},
	volume = {109},
	url = {https://link.aps.org/doi/10.1103/PhysRevB.109.085143},
	doi = {10.1103/PhysRevB.109.085143},
	number = {8},
	urldate = {2024-10-29},
	journal = {Physical Review B},
	author = {Song, Xue-Yang and Zhang, Ya-Hui and Senthil, T.},
	month = feb,
	year = {2024},
	pages = {085143},
}

@article{ludwig_integer_1994,
	title = {Integer quantum {Hall} transition: {An} alternative approach and exact results},
	volume = {50},
	copyright = {http://link.aps.org/licenses/aps-default-license},
	issn = {0163-1829, 1095-3795},
	shorttitle = {Integer quantum {Hall} transition},
	url = {https://link.aps.org/doi/10.1103/PhysRevB.50.7526},
	doi = {10.1103/PhysRevB.50.7526},
	language = {en},
	number = {11},
	urldate = {2025-06-28},
	journal = {Physical Review B},
	author = {Ludwig, Andreas W. W. and Fisher, Matthew P. A. and Shankar, R. and Grinstein, G.},
	month = sep,
	year = {1994},
	pages = {7526--7552},
}

@article{ye_coulomb_1998,
	title = {Coulomb {Interactions} at {Quantum} {Hall} {Critical} {Points} of {Systems} in a {Periodic} {Potential}},
	volume = {80},
	copyright = {http://link.aps.org/licenses/aps-default-license},
	issn = {0031-9007, 1079-7114},
	url = {https://link.aps.org/doi/10.1103/PhysRevLett.80.5409},
	doi = {10.1103/PhysRevLett.80.5409},
	language = {en},
	number = {24},
	urldate = {2024-12-09},
	journal = {Physical Review Letters},
	author = {Ye, Jinwu and Sachdev, Subir},
	month = jun,
	year = {1998},
	pages = {5409--5412},
}

@article{grover_entanglement_2014,
	title = {Entanglement {Monotonicity} and the {Stability} of {Gauge} {Theories} in {Three} {Spacetime} {Dimensions}},
	volume = {112},
	copyright = {http://link.aps.org/licenses/aps-default-license},
	issn = {0031-9007, 1079-7114},
	url = {https://link.aps.org/doi/10.1103/PhysRevLett.112.151601},
	doi = {10.1103/PhysRevLett.112.151601},
	language = {en},
	number = {15},
	urldate = {2024-11-20},
	journal = {Physical Review Letters},
	author = {Grover, Tarun},
	month = apr,
	year = {2014},
	pages = {151601},
}

@article{grover_quantum_2013,
	title = {Quantum phase transition between integer quantum {Hall} states of bosons},
	volume = {87},
	copyright = {http://link.aps.org/licenses/aps-default-license},
	issn = {1098-0121, 1550-235X},
	url = {https://link.aps.org/doi/10.1103/PhysRevB.87.045129},
	doi = {10.1103/PhysRevB.87.045129},
	language = {en},
	number = {4},
	urldate = {2024-09-23},
	journal = {Physical Review B},
	author = {Grover, Tarun and Vishwanath, Ashvin},
	month = jan,
	year = {2013},
	pages = {045129},
}

@article{lee_emergent_2018,
	title = {Emergent {Multi}-{Flavor} {QED} 3 at the {Plateau} {Transition} between {Fractional} {Chern} {Insulators}: {Applications} to {Graphene} {Heterostructures}},
	volume = {8},
	issn = {2160-3308},
	shorttitle = {Emergent {Multi}-{Flavor} {QED} 3 at the {Plateau} {Transition} between {Fractional} {Chern} {Insulators}},
	url = {https://link.aps.org/doi/10.1103/PhysRevX.8.031015},
	doi = {10.1103/PhysRevX.8.031015},
	language = {en},
	number = {3},
	urldate = {2025-01-26},
	journal = {Physical Review X},
	author = {Lee, Jong Yeon and Wang, Chong and Zaletel, Michael P. and Vishwanath, Ashvin and He, Yin-Chen},
	month = jul,
	year = {2018},
	pages = {031015},
}

@article{song_deconfined_2023,
	title = {Deconfined criticalities and dualities between chiral spin liquid, topological superconductor and charge density wave {Chern} insulator},
	volume = {15},
	issn = {2542-4653},
	url = {https://scipost.org/10.21468/SciPostPhys.15.5.215},
	doi = {10.21468/SciPostPhys.15.5.215},
	language = {en},
	number = {5},
	urldate = {2025-02-13},
	journal = {SciPost Physics},
	author = {Song, Xue-Yang and Zhang, Ya-Hui},
	month = nov,
	year = {2023},
	pages = {215},
}

@article{chen_mott_1993,
	title = {Mott transition in an anyon gas},
	volume = {48},
	url = {https://link.aps.org/doi/10.1103/PhysRevB.48.13749},
	doi = {10.1103/PhysRevB.48.13749},
	number = {18},
	urldate = {2025-02-02},
	journal = {Physical Review B},
	author = {Chen, Wei and Fisher, Matthew P. A. and Wu, Yong-Shi},
	month = nov,
	year = {1993},
	pages = {13749--13761},
}

@article{halperin_theory_1993,
	title = {Theory of the half-filled {Landau} level},
	volume = {47},
	url = {https://link.aps.org/doi/10.1103/PhysRevB.47.7312},
	doi = {10.1103/PhysRevB.47.7312},
	number = {12},
	urldate = {2024-09-01},
	journal = {Physical Review B},
	author = {Halperin, B. I. and Lee, Patrick A. and Read, Nicholas},
	month = mar,
	year = {1993},
	pages = {7312--7343},
}

@article{ma_emergent_2020,
	title = {Emergent \$\{{\textbackslash}mathrm\{{QCD}\}\}\_\{3\}\$ quantum phase transitions of fractional {Chern} insulators},
	volume = {2},
	url = {https://link.aps.org/doi/10.1103/PhysRevResearch.2.033348},
	doi = {10.1103/PhysRevResearch.2.033348},
	number = {3},
	urldate = {2025-02-13},
	journal = {Physical Review Research},
	author = {Ma, Ruochen and He, Yin-Chen},
	month = sep,
	year = {2020},
	pages = {033348},
}

@article{lu_extended_2025,
	title = {Extended quantum anomalous {Hall} states in graphene/{hBN} moiré superlattices},
	copyright = {2025 The Author(s), under exclusive licence to Springer Nature Limited},
	issn = {1476-4687},
	url = {https://www.nature.com/articles/s41586-024-08470-1},
	doi = {10.1038/s41586-024-08470-1},
	language = {en},
	urldate = {2025-01-27},
	journal = {Nature},
	author = {Lu, Zhengguang and Han, Tonghang and Yao, Yuxuan and Hadjri, Zach and Yang, Jixiang and Seo, Junseok and Shi, Lihan and Ye, Shenyong and Watanabe, Kenji and Taniguchi, Takashi and Ju, Long},
	month = jan,
	year = {2025},
	pages = {1--6},
}

@article{barkeshli_continuous_2014,
	title = {Continuous transition between fractional quantum {Hall} and superfluid states},
	volume = {89},
	copyright = {http://link.aps.org/licenses/aps-default-license},
	issn = {1098-0121, 1550-235X},
	url = {https://link.aps.org/doi/10.1103/PhysRevB.89.235116},
	doi = {10.1103/PhysRevB.89.235116},
	language = {en},
	number = {23},
	urldate = {2025-03-24},
	journal = {Physical Review B},
	author = {Barkeshli, Maissam and McGreevy, John},
	month = jun,
	year = {2014},
	pages = {235116},
}

@article{barkeshli_continuous_2012,
	title = {Continuous transitions between composite {Fermi} liquid and {Landau} {Fermi} liquid: {A} route to fractionalized {Mott} insulators},
	volume = {86},
	copyright = {http://link.aps.org/licenses/aps-default-license},
	issn = {1098-0121, 1550-235X},
	shorttitle = {Continuous transitions between composite {Fermi} liquid and {Landau} {Fermi} liquid},
	url = {https://link.aps.org/doi/10.1103/PhysRevB.86.075136},
	doi = {10.1103/PhysRevB.86.075136},
	language = {en},
	number = {7},
	urldate = {2024-12-11},
	journal = {Physical Review B},
	author = {Barkeshli, Maissam and McGreevy, John},
	month = aug,
	year = {2012},
	pages = {075136},
}

@article{barkeshli_anyon_2010,
	title = {Anyon {Condensation} and {Continuous} {Topological} {Phase} {Transitions} in {Non}-{Abelian} {Fractional} {Quantum} {Hall} {States}},
	volume = {105},
	copyright = {http://link.aps.org/licenses/aps-default-license},
	issn = {0031-9007, 1079-7114},
	url = {https://link.aps.org/doi/10.1103/PhysRevLett.105.216804},
	doi = {10.1103/PhysRevLett.105.216804},
	language = {en},
	number = {21},
	urldate = {2024-08-11},
	journal = {Physical Review Letters},
	author = {Barkeshli, Maissam and Wen, Xiao-Gang},
	month = nov,
	year = {2010},
	pages = {216804},
}

@article{wen_transitions_1993,
	title = {Transitions between the quantum {Hall} states and insulators induced by periodic potentials},
	volume = {70},
	copyright = {http://link.aps.org/licenses/aps-default-license},
	issn = {0031-9007},
	url = {https://link.aps.org/doi/10.1103/PhysRevLett.70.1501},
	doi = {10.1103/PhysRevLett.70.1501},
	language = {en},
	number = {10},
	urldate = {2024-08-11},
	journal = {Physical Review Letters},
	author = {Wen, Xiao-Gang and Wu, Yong-Shi},
	month = mar,
	year = {1993},
	pages = {1501--1504},
}

@article{kuhlenkamp_chiral_2024,
  title = {Chiral Pseudospin Liquids in Moir\'e Heterostructures},
  author = {Kuhlenkamp, Clemens and Kadow, Wilhelm and {\.I}mamo{\u g}lu, Ata{\c c} and Knap, Michael},
  journal = {Phys. Rev. X},
  volume = {14},
  issue = {2},
  pages = {021013},
  numpages = {13},
  year = {2024},
  month = {Apr},
  publisher = {American Physical Society},
  doi = {10.1103/PhysRevX.14.021013},
  url = {https://link.aps.org/doi/10.1103/PhysRevX.14.021013}
}

@article{divic_anyon_2025,
author = {Stefan Divic  and Valentin Crépel  and Tomohiro Soejima  and Xue-Yang Song  and Andrew J. Millis  and Michael P. Zaletel  and Ashvin Vishwanath },
title = {Anyon superconductivity from topological criticality in a Hofstadter–Hubbard model},
journal = {Proceedings of the National Academy of Sciences},
volume = {122},
number = {33},
pages = {e2426680122},
year = {2025},
doi = {10.1073/pnas.2426680122},
URL = {https://www.pnas.org/doi/abs/10.1073/pnas.2426680122},
eprint = {https://www.pnas.org/doi/pdf/10.1073/pnas.2426680122},
}

@misc{divic_chiral_2025,
      title={Chiral Spin Liquid and Quantum Phase Transition in the Triangular Lattice Hofstadter-Hubbard Model}, 
      author={Stefan Divic and Tomohiro Soejima and Valentin Crépel and Michael P. Zaletel and Andrew Millis},
      year={2025},
      eprint={2406.15348},
      archivePrefix={arXiv},
      primaryClass={cond-mat.str-el},
      url={https://arxiv.org/abs/2406.15348}, 
}

@misc{kuhlenkamp_robust_2025,
      title={Robust superconductivity upon doping chiral spin liquid and Chern insulators in a Hubbard-Hofstadter model}, 
      author={Clemens Kuhlenkamp and Stefan Divic and Michael P. Zaletel and Tomohiro Soejima and Ashvin Vishwanath},
      year={2025},
      eprint={2509.02675},
      archivePrefix={arXiv},
      primaryClass={cond-mat.str-el},
      url={https://arxiv.org/abs/2509.02675}, 
}

@misc{chen_topological_2025,
      title={Topological Chiral Superconductivity in the Triangular-Lattice Hofstadter-Hubbard Model}, 
      author={Feng Chen and Wen O. Wang and Jia-Xin Zhang and Leon Balents and D. N. Sheng},
      year={2025},
      eprint={2509.02757},
      archivePrefix={arXiv},
      primaryClass={cond-mat.str-el},
      url={https://arxiv.org/abs/2509.02757}, 
}

@misc{kousa_theory_2025,
      title={Theory of magnetoroton bands in moir\'e materials}, 
      author={Bishoy M. Kousa and Nicolás Morales-Durán and Tobias M. R. Wolf and Eslam Khalaf and Allan H. MacDonald},
      year={2025},
      eprint={2502.17574},
      archivePrefix={arXiv},
      primaryClass={cond-mat.mes-hall},
      url={https://arxiv.org/abs/2502.17574}, 
}

@article{Wu_moire_2016,
  title = {Moir\'e assisted fractional quantum Hall state spectroscopy},
  author = {Wu, Fengcheng and MacDonald, A. H.},
  journal = {Phys. Rev. B},
  volume = {94},
  issue = {24},
  pages = {241108},
  numpages = {5},
  year = {2016},
  month = {Dec},
  publisher = {American Physical Society},
  doi = {10.1103/PhysRevB.94.241108},
  url = {https://link.aps.org/doi/10.1103/PhysRevB.94.241108}
}

@misc{zhang_pathways_2025,
      title={Pathways from a chiral superconductor to a composite Fermi liquid}, 
      author={Yunchao Zhang and Leyna Shackleton and T. Senthil},
      year={2025},
      eprint={2509.21591},
      archivePrefix={arXiv},
      primaryClass={cond-mat.str-el},
      url={https://arxiv.org/abs/2509.21591}, 
}

@article{Goldman_zerofield_2023,
  title = {Zero-Field Composite Fermi Liquid in Twisted Semiconductor Bilayers},
  author = {Goldman, Hart and Reddy, Aidan P. and Paul, Nisarga and Fu, Liang},
  journal = {Phys. Rev. Lett.},
  volume = {131},
  issue = {13},
  pages = {136501},
  numpages = {8},
  year = {2023},
  month = {Sep},
  publisher = {American Physical Society},
  doi = {10.1103/PhysRevLett.131.136501},
  url = {https://link.aps.org/doi/10.1103/PhysRevLett.131.136501}
}

@article{Hafezi_fractional_2007,
  title = {Fractional quantum Hall effect in optical lattices},
  author = {Hafezi, M. and S\o{}rensen, A. S. and Demler, E. and Lukin, M. D.},
  journal = {Phys. Rev. A},
  volume = {76},
  issue = {2},
  pages = {023613},
  numpages = {16},
  year = {2007},
  month = {Aug},
  publisher = {American Physical Society},
  doi = {10.1103/PhysRevA.76.023613},
  url = {https://link.aps.org/doi/10.1103/PhysRevA.76.023613}
}

@misc{liu_fractional_2025,
      title={From fractional Chern insulators to topological electronic crystals in moir\'e {MoTe}$_2$: quantum geometry tuning via remote layer}, 
      author={Feng Liu and Fan Xu and Cheng Xu and Jiayi Li and Zheng Sun and Jiayong Xiao and Ning Mao and Xumin Chang and Xinglin Tao and Kenji Watanabe and Takashi Taniguchi and Jinfeng Jia and Ruidan Zhong and Zhiwen Shi and Shiyong Wang and Guorui Chen and Xiaoxue Liu and Dong Qian and Yang Zhang and Tingxin Li and Shengwei Jiang},
      year={2025},
      eprint={2512.03622},
      archivePrefix={arXiv},
      primaryClass={cond-mat.mes-hall},
      url={https://arxiv.org/abs/2512.03622}, 
}

@misc{pierce_imaging_2025,
      title={Imaging propagating terahertz collective modes in two-dimensional semiconductor double layers}, 
      author={Andrew T. Pierce and Chirag Vaswani and Dimitri Pimenov and Sihong Xu and Kenji Watanabe and Takashi Taniguchi and Erich Mueller and Debanjan Chowdhury and Kin Fai Mak and Jie Shan},
      year={2025},
      eprint={2511.22962},
      archivePrefix={arXiv},
      primaryClass={cond-mat.mes-hall},
      url={https://arxiv.org/abs/2511.22962}, 
}

@article{chen_direct_2025,
  title = {Direct {{Measurement}} of {{Terahertz Conductivity}} in a {{Gated Monolayer Semiconductor}}},
  author = {Chen, Su-Di and Feng, Qixin and Zhao, Wenyu and Qi, Ruishi and Zhang, Zuocheng and Abeysinghe, Dishan and Uzundal, Can and Xie, Jingxu and Taniguchi, Takashi and Watanabe, Kenji and Wang, Feng},
  year = 2025,
  month = may,
  journal = {Nano Letters},
  volume = {25},
  number = {19},
  pages = {7998--8002},
  doi = {10.1021/acs.nanolett.5c01605},
  pmid = {40323824}
}

@misc{xu_plasmon_2026,
      title={Plasmon dynamics in graphene}, 
      author={Suheng Xu and Birui Yang and Nishchhal Verma and Rocco A. Vitalone and Brian Vermilyea and Miguel Sánchez Sánchez and Julian Ingham and Ran Jing and Yinming Shao and Tobias Stauber and Angel Rubio and Milan Delor and Mengkun Liu and Michael M. Fogler and Cory R. Dean and Andrew Millis and Raquel Queiroz and D. N. Basov},
      year={2026},
      eprint={2601.10493},
      archivePrefix={arXiv},
      primaryClass={cond-mat.mes-hall},
      url={https://arxiv.org/abs/2601.10493}, 
}

@article{xu_electronic_2024,
author = {Suheng Xu  and Yutao Li  and Rocco A. Vitalone  and Ran Jing  and Aaron J. Sternbach  and Shuai Zhang  and Julian Ingham  and Milan Delor  and James W. McIver  and Matthew Yankowitz  and Raquel Queiroz  and Andrew J. Millis  and Michael M. Fogler  and Cory R. Dean  and Abhay N. Pasupathy  and James Hone  and Mengkun Liu  and D. N. Basov },
title = {Electronic interactions in Dirac fluids visualized by nano-terahertz spacetime interference of electron-photon quasiparticles},
journal = {Science Advances},
volume = {10},
number = {43},
pages = {eado5553},
year = {2024},
doi = {10.1126/sciadv.ado5553},
URL = {https://www.science.org/doi/abs/10.1126/sciadv.ado5553},
eprint = {https://www.science.org/doi/pdf/10.1126/sciadv.ado5553},
}

\end{document}